\documentclass[12pt,prd,tightenlines,nofootinbib,showpacsshowkeys]{revtex4}
\newcommand{\be}{\begin{equation}}
\newcommand{\ee}{\end{equation}}
\newcommand{\bga}{\begin{equation}\begin{gathered}}
\newcommand{\ega}{\end{gathered}\end{equation}}
\newcommand{\bgas}{\begin{equation}\begin{gathered}}
\newcommand{\egas}{\end{gathered}\end{equation}}

\usepackage{bm}
\usepackage{graphics}
\usepackage{rotating}
\usepackage{epsfig}
\usepackage{amsmath}
\usepackage{amsfonts}

\begin{document}
\title{\begin{flushright}{\rm\normalsize SSU-HEP-14/11\\[5mm]}\end{flushright}
Ground state hyperfine structure in muonic lithium ions}
\author{\firstname{A.~P.} \surname{Martynenko}}
\affiliation{Samara State University, Pavlov Str. 1, 443011, Samara, Russia}
\affiliation{Samara State Aerospace University named after S.P. Korolyov, Moskovskoye Shosse 34, 443086,
Samara, Russia}
\author{\firstname{A.~A.} \surname{Ulybin}}
\affiliation{Samara State University, Pavlov Str. 1, 443011, Samara, Russia }

\begin{abstract}
On the basis of perturbation theory in fine structure constant
$\alpha$ and the ratio of electron to muon masses we calculate
one-loop vacuum polarization, electron vertex corrections, nuclear
structure and recoil corrections to hyperfine splitting of the
ground state in muonic lithium ions $(\mu\ e\ ^6_3Li)^+$ and $(\mu\
e\ ^7_3Li)^+$. We obtain total results for the ground state
small hyperfine splittings in $(\mu\ e\ ^6_3Li)^+$ $\Delta\nu_1=21572.16$ MHz
and $\Delta\nu_2=14152.56$ MHz and in $(\mu\ e\ ^7_3Li)^+$ $\Delta\nu_1=21733.06$ MHz
and $\Delta\nu_2=13994.35$ MHz which
can be considered as a reliable estimate for a comparison with
future experimental data.
\end{abstract}


\maketitle

\section{Introduction}

Muonic lithium ions $(\mu\ e\  ^6_3Li)^+$ and $(\mu\ e\  ^7_3Li)^+$
represent simple three-body atomic systems consisting of one electron, negative
charged muon and positive charged nucleus $^6_3Li$ or $^7_3Li$. A lifetime of
muonic atoms is determined by muon lifetime $\tau_\mu=2.19703(4)\cdot 10^{-6}$ s.
It is greater than the time of atomic processes, so, the muon has a time to make a number
of transitions between energy levels which are attended by $\gamma$-radiation.
These three-particle systems have complicate ground state hyperfine structure which appears
due to the interaction of magnetic moments of the electron, muon and nucleus.
Light muonic atoms represent a unique laboratory for precise determination of nuclear properties
such as the nuclear charge radius \cite{CREMA,CREMA1}. In last years we observe an essential progress
achieved by the CREMA (Charge Radius Experiment with Muonic Atoms) collaboration in the study of the
energy structure in muonic hydrogen. The measurement of the Lamb shift $(2P-2S)$ and hyperfine splitting
of $2S$-state allows to find more precise value of proton charge radius $r_p= 0.84087(39)$ fm, the Zemach
radius $r_Z = 1.082(37)$ fm and magnetic radius $r_M = 0.87(6)$ fm.
The obtained value of proton charge radius $r_p$ is an order of magnitude
more precise than the 2010-CODATA value which was derived by means of
different methods including the hydrogen spectroscopy. It differs from CODATA value
by $7\sigma$.
Note that the Zemach radius of the proton $r_Z = 1.045(16)$ fm
and magnetic radius $r_M = 0.778(29)$ fm were obtained earlier more precisely
from a comparison of experimental data with predictions for the hydrogen hyperfine
splitting in \cite{volotka}. Similar measurements are performed also
in the case of muonic deuterium and ions of muonic helium and planned for a publication.
Light muonic
atoms are important for the check of Standard Model and bound state theory in quantum electrodynamics,
for search of exotic interactions of elementary particles. Thus, for example, muonic systems can be used
for the search of Lorentz and CPT symmetry violations \cite{drake}.

Hyperfine splitting (HFS) of the ground state of muonic helium atom $(\mu\ e\ ^3_2He)$ was measured many
years ago with sufficiently high accuracy in \cite{Gladish}. This is the only experiment with muonic
three-particle systems. In turn theoretical investigations of the energy spectrum of muonic helium atom
and other three-particle systems
achieved much successes in two directions \cite{LM,HH,Borie,RD1,Chen,Amusia,AF,AF1}.
First approach in \cite{LM,HH,Amusia} is based on the perturbation theory (PT) for
the Schr\"odinger equation. In this case there is analytical solution for three-particle
bound state wave function in initial approximation. Using it a calculation of
different corrections to HFS can be performed. Another approach in \cite{RD1,AF,Chen,VK2000,KP2001}
is built on the variational method which allows to calculate numerically the energy levels in three-particle
systems with high accuracy. In order to find an arrangement of the low lying energy levels with high precision
we should take into account different corrections to particle interaction operator.
First of all these corrections are related with recoil effects, vacuum polarization and nuclear structure effects.
The program of analytical calculation of hyperfine splitting in muonic helium atom including excited state
was realized in \cite{LM,HH,Chen,Amusia,AF,km}. It allowed to present hyperfine splitting in analytical 
form as a series into small parameters existing in this task.
In this work we aim to extend that approach on muonic lithium ions which
represent potentially an interest for experimental study. So, the purpose of this paper is to provide a detailed
calculation of hyperfine splittings for the systems $(\mu\ e\ ^6_3Li)^+$, $(\mu\ e\ ^7_3Li)^+$.

The bound particles in muonic lithium ions have different masses $m_e\ll m_\mu\ll m_{Li}$. As a result
the muon and Li nucleus compose the pseudonucleus $(\mu\ ^{6,7}_3Li)^{++}$ and muonic lithium ion looks
as a two-particle system in the first approximation. Three-particle bound system $(\mu\ e\  ^{6,7}_3Li)^+$ is described
by the Hamiltonian:
\begin{equation}
\label{eq:1}
H=H_0+\Delta H+\Delta H_{rec}+\Delta H_{VP}+\Delta H_{str}+\Delta H_{vert},~H_0=-\frac{1}{2M_\mu}\nabla^2_\mu-\frac{1}{2M_e}
\nabla^2_e-\frac{3\alpha}{x_\mu}-\frac{2\alpha}{x_e},
\end{equation}
\begin{equation}
\label{eq:2}
\Delta H=\frac{\alpha}{x_{\mu e}}-\frac{\alpha}{x_e},~~~\Delta
H_{rec}=-\frac{1}{M_{Li}}
{\mathstrut\bm\nabla}_\mu\cdot{\mathstrut\bm\nabla}_e,
\end{equation}
where ${\bf x_\mu}$ and ${\bf x_e}$ are the muon and electron coordinates
relative to the lithium nucleus, $M_e=m_eM_{Li}/(m_e+M_{Li})$,
$M_\mu=m_\mu M_{Li}/(m_\mu+M_{Li})$ are the reduced masses of subsystems
$(e\ {^{6,7}_3Li})^{++}$ and $(\mu\ ^{6,7}_3Li)^{++}$. The Hamiltonian terms $\Delta H_{VP}$,
$\Delta H_{str}$ and $\Delta H_{vert}$ which describe vacuum polarization, structure and vertex
corrections are constructed below. In initial approximation the wave function of the ground state has the form:
\begin{equation}
\label{eq:3}
\Psi_0({\bf x_e},{\bf x_\mu})=\psi_{e0}({\bf x_e})\psi_{\mu 0}({\bf
x_\mu})=\frac{1}{\pi} (6\alpha^2M_eM_\mu)^{3/2}e^{-3\alpha M_\mu
x_\mu}e^{-2\alpha M_e x_e}.
\end{equation}

As it follows from the structure of the Hamiltonian presented in
\eqref{eq:1}-\eqref{eq:2} we include in basic Hamiltonian $H_0$ only a part of the Coulomb electron-nucleus
interaction. The reminder is considered as a perturbation as well as the Coulomb muon-electron
interaction. In this way we can explore analytical method of the calculation of hyperfine
structure based on the perturbation theory. Analytical solution for the wave function \eqref{eq:3}
allows to obtain
the perturbation contributions in two small parameters $\alpha$ and $M_e/M_\mu$ as demonstrated
below. The corrections due to electron-muon interaction and mass polarization term (2) are
considered in second orders of perturbation theory in subsequent sections.
They are presented firstly in analytical integral form and calculated after that analytically
or numerically.

Basic contribution to hyperfine interaction in the ground state of $(\mu\ e\ ^{6,7}_3Li)^+$
is determined by the following Hamiltonian:
\begin{equation}
\label{eq:4}
\Delta H_0^{hfs}=\frac{2\pi\alpha}{3}\frac{g_\mu g_N}{m_\mu m_p}({\bf S}_\mu\cdot{\bf I})\delta({\bf
x}_\mu)-\frac{2\pi\alpha}{3}\frac{g_eg_\mu}{m_em_\mu}({\bf S}_e\cdot{\bf S}_\mu)\delta({\bf
x}_\mu-{\bf
x}_e)+\frac{2\pi\alpha}{3}\frac{g_eg_N}{m_em_p}({\bf S}_e\cdot{\bf I})\delta({\bf x}_e),
\end{equation}
where $g_e$, $g_\mu$ and $g_N$ are gyromagnetic factors of the electron, muon and nucleus.
Total spin of
three spin particles can be either 2, 1 and 0 for $(\mu\ e\ ^{6}_3Li)^+$ and
5/2, 3/2 and 1/2 for $(\mu\ e\ ^{7}_3Li)^+$.

\begin{figure}
\centering
\includegraphics[width=7.cm]{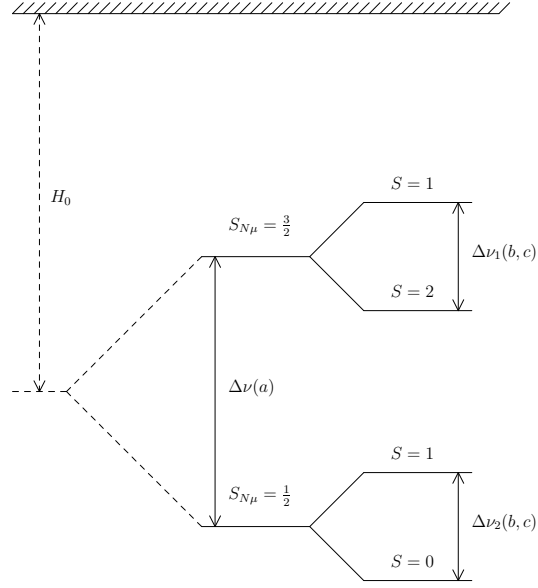}
\caption{Schematic hypefine splittings of the ground state in muonic
lithium ions. Numerical values of angular momenta are presented in the case of $(\mu\ e\ ^6_3Li)^+$.}
\label{fig:fig_Li}
\end{figure}

Hyperfine splitting of the energy levels in muonic lithium ions is determined by the following matrix elements:
\begin{equation}
\label{eq:5}
\nu=\langle \Delta H^{hfs}_0\rangle=a\ \langle {\bf I}\cdot{\bf S}_\mu\rangle-b\ \langle {\bf
S}_\mu\cdot{\bf S}_e\rangle +c\ \langle {\bf S}_e\cdot{\bf I}\rangle,
\end{equation}
where the spin-space expectation values can be calculated using the following basis
transformation \cite{iis}:
\begin{equation}
\label{eq:6}
\Psi_{S_{N\mu}SS_z}=\sum_{S_{Ne}} (-1)^{S_\mu+I+S_e+S}\sqrt{(2S_{N\mu}+1)(2S_{Ne}+1)}
\left\{\begin{array}{ccc}
S_e& S_N & S_{Ne} \\
S_\mu & S & S_{N\mu} \\
\end{array}\right\}\Psi_{S_{Ne}SS_z}.
\end{equation}
$S_{N\mu}$ is the spin in muon-nucleus subsystem, $S_{Ne}$ is the spin in electron-nucleus subsystem,
$S$ is total angular momentum. The properties of $6j$-symbols also can be found in \cite{iis}.
As it follows from \eqref{eq:4}-\eqref{eq:5} basic contributions to coefficients $a$, $b$ and $c$ are the following:
\begin{equation}
\label{eq:7}
a_0=\frac{2\pi\alpha}{3}\frac{g_Ng_\mu}{m_pm_\mu}\langle \delta({\bf
x}_\mu)\rangle,~b_0=\frac{2\pi\alpha}{3}\frac{g_\mu g_e}{m_\mu
m_e}\langle \delta({\bf x}_\mu-{\bf
x}_e)\rangle,~c_0=\frac{2\pi\alpha}{3}\frac{g_eg_N}{m_em_p}\langle\delta({\bf
x}_e)\rangle,
\end{equation}
where $\langle...\rangle$ denotes the expectation value in coordinate space over wave functions \eqref{eq:3}.
We have to take into account numerical values of gyromagnetic factors $g_e=2$ for the $b$ coefficient,
$g_e=2(1+\kappa_e)=2(1+ 1.15965218111(74)\cdot 10^{-3})$ for the $c$ coefficient,
$g_\mu=2(1+\kappa_\mu)=2\cdot (1+1.16592069(60)\cdot 10^{-3})$,
$g_N(^6_3Li)=0.822047$, $g_N(^7_3Li)=2.170951$.

The expectation value \eqref{eq:5} is the $4\times 4$ matrix corresponding to different values
of total spin and muon-nucleus spin: ($S=0, S_{N\mu}=\frac{1}{2}$), ($S=1, S_{N\mu}=\frac{1}{2}$),
($S=1, S_{N\mu}=\frac{3}{2}$), ($S=2, S_{N\mu}=\frac{3}{2}$) for the ion $(\mu\ e\ ^6_3Li)^+$;
($S=\frac{1}{2}, S_{N\mu}=1$), ($S=\frac{3}{2}, S_{N\mu}=1$),
($S=\frac{3}{2}, S_{N\mu}=2$), ($S=\frac{5}{2}, S_{N\mu}=2$) for the ion $(\mu\ e\ ^7_3Li)^+$.
After its diagonalization
we obtain four energy eigenvalues $\nu_i$. In the case of muonic
lithium ions we have relations $a\gg b$ and $a\gg c$. So, small
hyperfine splitting intervals $\Delta\nu_i$ related to the experiment
can be written with good accuracy in simple form:
\begin{equation}
\label{eq:8}
\Delta\nu_1^{hfs}(\mu\ e\ ^6_3Li)=\frac{2(b-2c)}{3}+O\left(\frac{b}{a},\frac{c}{a}\right),~
\Delta\nu_2^{hfs}(\mu\ e\ ^6_3Li)=\frac{b+4c}{3}+O\left(\frac{b}{a},\frac{c}{a}\right),
\end{equation}
\begin{equation}
\label{eq:9}
\Delta\nu_1^{hfs}(\mu\ e\ ^7_3Li)=\frac{5(b-3c)}{8}+O\left(\frac{b}{a},\frac{c}{a}\right),~
\Delta\nu_2^{hfs}(\mu\ e\ ^7_3Li)=\frac{3(b+5c)}{8}+O\left(\frac{b}{a},\frac{c}{a}\right).
\end{equation}

For angular momentum of muon-nucleus subsystem $S_{\mu N}=3/2$ and
$S_{\mu N}=1/2$ $(\mu \ e\ ^6_3Li)^+$ hyperfine splitting intervals
\eqref{eq:8} between states with total angular momentum $S=2,1$ and
$S=1,0$ arise from magnetic interaction between the electron and
pseudonucleus $(\mu\ ^6_3Li)^{++}$. The same situation is valid for
hyperfine splitting intervals \eqref{eq:9} for $(\mu \ e\
^7_3Li)^+$. Schematic diagram of hyperfine splittings in muonic
lithium ions is presented in Fig.~\ref{fig:fig_Li}

In first order perturbation theory (PT) basic contributions to the coefficients $b$ and $c$ \eqref{eq:7}
can be calculated analytically using \eqref{eq:3} (hereinafter the upper and lower values correspond to $(\mu\ e\ ^6_3Li)^+$
and $(\mu\ e\ ^7_3Li)^+$):
\begin{equation}
\label{eq:10}
b_0=\frac{2\pi \alpha}{3}\frac{g_eg_\mu}{m_em_\mu}\int\Psi^\ast({\bf x}_e,{\bf x}_\mu)\delta({\bf x}_e-{\bf x}_\mu)\Psi({\bf x}_e,{\bf x}_\mu)
d{\bf x}_ed{\bf x}_\mu=\nu_F\frac{g_eg_\mu}{4}\frac{1}{\left(1+\frac{2M_e}{3M_\mu}\right)^3}=
\end{equation}
\begin{displaymath}
=\nu_F\left[1+\kappa_\mu+(1+\kappa_\mu)\left(-2\frac{M_e}{M_\mu}+\frac{8}{3}\frac{M_e^2}{M_\mu^2}\right)\right],
~\nu_F=\frac{64M_e^3\alpha^4}{3m_em_\mu}=\Biggl\{{{36140.290~MHz}\atop{36141.701~MHz}},
\end{displaymath}
\begin{equation}
\label{eq:11}
c_0=\frac{2\pi \alpha}{3}\frac{g_eg_N}{m_em_p}\int\Psi^\ast({\bf x}_e,{\bf x}_\mu)\delta({\bf x}_e)\Psi({\bf x}_e,{\bf x}_\mu)
d{\bf x}_ed{\bf x}_\mu=\nu_F\frac{m_\mu}{m_p}\frac{g_eg_N}{4}=\Biggl\{{{1674.700~MHz}\atop{4422.900~MHz}},
\end{equation}
where we have extracted in square brackets the Fermi energy $\nu_F$, muon anomalous magnetic moment correction
$\kappa_\mu\nu_F$ and recoil terms. Their corresponding numerical values for two lithium ions are presented in Table~\ref{tb1}.

Note, that, as we determine contributions to the energy spectrum numerically, corresponding results are presented with an
accuracy of 0.001 MHz. We express further the hyperfine splitting contributions in the frequency unit using the relation $\Delta
E^{hfs}=2\pi\hbar\Delta \nu^{hfs}$.
Modern numerical values of fundamental physical constants
are taken from~\cite{MT,sgk,stone}: the electron mass $m_e=0.510998928(11)\cdot
10^{-3}$ GeV, the muon mass $m_\mu=0.1056583715(35)$ GeV, fine structure constant
$\alpha^{-1}=137.035999074(44)$, the proton mass $m_p$ = 0.938272046(21)~GeV,
magnetic moments of Li nucleus in nuclear magnetons: $\mu(^6_3Li)$=0.8220473(6),
$\mu(^7_3Li)$=3.256427(2), masses of Li nucleus $M(^6_3Li)$=5.60152~GeV, $M(^7_3Li)$=6.53383~GeV
muon anomalous magnetic moment $\kappa_\mu=1.16592091(63)\cdot 10^{-3}$, the electron anomalous
magnetic moment $\kappa_e= 1.15965218076(27)\cdot 10^{-3}$.

In next sections we calculate different corrections to coefficients $b_0$ and $c_0$ over two small
parameters $\alpha$ and $M_e/M_\mu$.

\begin{figure}[htbp]
\centering
\includegraphics{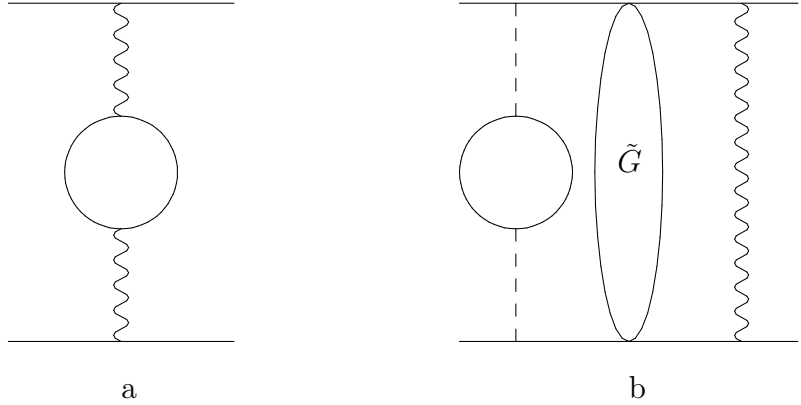}
\caption{The vacuum polarization effects. The dashed line represents
the Coulomb photon. The wave line represents the hyperfine part of
the Breit potential. $\tilde G$ is the reduced Coulomb Green's
function.}
\label{fig:pic1}
\end{figure}

\section{Recoil corrections}

Let us consider a calculation of recoil corrections of order $\alpha^4\frac{M_e}{M_\mu}$,
$\alpha^4\frac{M^2_e}{M^2_\mu}\ln\frac{M_e}{M_\mu}$ and $\alpha^4\frac{M^2_e}{M^2_\mu}$.
Using basic relations obtained in \cite{LM} for muonic helium atom we present here corresponding
results for muonic lithium ions.
A part of such corrections appears already in \eqref{eq:10}. In second order PT we also have the
contribution to hyperfine splitting which contains necessary order corrections. The correction to
the coefficient $b$ is the following:
\begin{equation}
\label{eq:12}
b_1=2\int\Psi^\ast({\bf x}_e,{\bf x}_\mu)\Delta H_0^{hfs}({\bf x}_e-{\bf x}_\mu)\tilde G({\bf x}_e,{\bf x}_\mu;{\bf x'}_e,{\bf x'}_\mu)
\Delta H({\bf x'}_e,{\bf x'}_\mu)\Psi({\bf x'}_e,{\bf x'}_\mu)d{\bf x}_ed{\bf x}_\mu d{\bf x'}_ed{\bf x'}_\mu,
\end{equation}
where the reduced Coulomb Green's function has the form:
\begin{equation}
\label{eq:13}
\tilde G({\bf x}_e,{\bf x}_\mu;{\bf x'}_e,{\bf x'}_\mu)=\sum_{n,n'\not =0}\frac{\psi_{\mu n}({\bf x}_\mu)
\psi_{en'}({\bf x}_e)\psi^\ast_{\mu n}({\bf x'}_\mu)\psi^\ast_{en'}({\bf x'}_e)}{E_{\mu 0}+E_{e0}-E_{\mu n}-E_{en'}}.
\end{equation}
Dividing the sum over muon states into two parts with $n=0$ and $n\not =0$ we obtain for the first part:
\begin{equation}
\label{eq:14}
b_1(n=0)=\frac{4\pi \alpha}{3}\frac{g_eg_\mu}{m_em_\mu}\int|\psi_{\mu 0}({\bf x}_3)|^2\psi^\ast_{e0}({\bf x}_3)\sum_{n'\not=0}^\infty
\frac{\psi_{en'}({\bf x}_3)\psi^\ast_{en'}({\bf x}_1)}{E_{e0}-E_{en'}}V_\mu({\bf x}_1)\psi_{e0}({\bf x}_1)d{\bf x}_1d{\bf x}_3,
\end{equation}
\begin{equation}
\label{eq:15}
V_\mu({\bf x}_1)=\int\psi^\ast_{\mu 0}({\bf x}_2)\left[\frac{\alpha}{|{\bf x}_2-{\bf x}_1|}-\frac{\alpha}{x_1}\right]\psi_{\mu 0}({\bf x}_2)d{\bf x}_2=
-\frac{\alpha}{x_1}(1+3\alpha x_1M_\mu)e^{-6\alpha x_1M_\mu}.
\end{equation}
For the further integration in \eqref{eq:14} over coordinates we use compact expression of the electron reduced Coulomb Green's function
obtained in \cite{hameka}:
\begin{equation}
\label{eq:16}
G_e({\bf x}_1,{\bf x}_3)=\sum_{n\not =0}^\infty\frac{\psi_{en}({\bf
x}_3) \psi_{en}^\ast({\bf x}_1)}{E_{e0}-E_{en}}=-\frac{2\alpha
M_e^2}{\pi}e^{-2\alpha M_e(x_1+x_3)}\Biggl[\frac{1}{4\alpha M_e x_>}-
\end{equation}
\begin{displaymath}
-\ln(4\alpha M_e x_>)-\ln(4\alpha M_e x_<)+Ei(4\alpha M_e x_<)+
\frac{7}{2}-2C-2\alpha M_e(x_1+x_3)+\frac{1-e^{4\alpha M_e
x_<}}{4\alpha M_e x_<}\Biggr],
\end{displaymath}
where $x_<=\min(x_1,x_3)$, $x_>=\max(x_1,x_3)$, $C=0.577216\ldots$
is the Euler's constant and $Ei(x)$ is the exponential-integral function.
The result of coordinate integration in \eqref{eq:14} can be written as an expansion in $M_e/M_\mu$:
\begin{equation}
\label{eq:17}
b_1(n=0)=\nu_F(1+\kappa_\mu)\left[\frac{11}{24}\frac{M_e}{M_\mu}+\frac{1}{72}\frac{M_e^2}{M_\mu^2}\left(
-64\ln\frac{M_e}{M_\mu}-7-128\ln 2+64\ln 3\right)\right].
\end{equation}
Second contribution to $b$ corresponding to muon excited states is equal to
\begin{equation}
\label{eq:18}
b_1(n\not=0)=\frac{4\pi\alpha}{3}\frac{g_eg_\mu}{m_em_\mu}\int\psi^\ast_{\mu 0}({\bf x}_3)\psi^\ast_{e 0}({\bf x}_3)
\sum_{n\not=0}\psi_{\mu n}({\bf x}_3)\psi^\ast_{\mu n}({\bf x}_2)G_e({\bf x}_3,{\bf x}_1,z)\times
\end{equation}
\begin{displaymath}
\left[\frac{\alpha}{|{\bf x}_2-{\bf x}_1|}-\frac{\alpha}{x_1}\right]\psi_{\mu 0}({\bf x}_2)\psi_{e 0}({\bf x}_1)d{\bf x}_1
d{\bf x}_2d{\bf x}_3,
\end{displaymath}
where the electron Coulomb Green's function
\begin{equation}
\label{eq:19}
G_e({\bf x}_3,{\bf x}_1,z)=\sum_{n'=0}^\infty\frac{\psi_{en'}({\bf x}_3)\psi^\ast_{en'}({\bf x}_1)}{z-E_{en'}}=
\sum_{n'=0}^\infty\frac{\psi_{en'}({\bf x}_3)\psi^\ast_{en'}({\bf x}_1)}{E_{\mu 0}+E_{e0}-E_{\mu n}-E_{en'}}.
\end{equation}
The term $(-\alpha/x_1)$ does not contribute due to the orthogonality of muon wave functions. In order to make analytical
integration in \eqref{eq:18} we use a replacement of $G_e$ by free electron Green's function \cite{LM}:
\begin{equation}
\label{eq:20}
G_e({\bf x}_3,{\bf x}_1,E_{\mu 0}+E_{e0}-E_{\mu n})\to G_{e0}({\bf x}_3-{\bf x}_1,E_{\mu 0}+E_{e0}-E_{\mu n})=-\frac{M_e}{2\pi}\frac{e^{-\beta|{\bf x}_3-{\bf x}_1|}}{|{\bf x}_3-{\bf x}_1|},
\end{equation}
where $\beta=\sqrt{2M_e(E_{\mu n}-E_{e0}-E_{\mu 0})}$. Moreover, we replace the electron wave functions in \eqref{eq:18} by their values
at the origin $\psi_{e0}(0)$. The omitted in this approximation terms can give contributions of second order in $\frac{M_e}{M_\mu}$.
The results of numerical integration presented in \cite{LM} for muonic helium show that these corrections are numerically small.
After used approximations an analytical integration over coordinate ${\bf x}_1$ gives the result:
\begin{equation}
\label{eq:21}
\int\frac{e^{-\beta|{\bf x}_3-{\bf x}_1|}}{|{\bf x}_3-{\bf x}_1|}\frac{d{\bf x}_1}{|{\bf x}_2-{\bf x}_1|}=
4\pi\left[\frac{1}{\beta}-\frac{1}{2}|{\bf x}_3-{\bf x}_2|+\frac{1}{6}\beta|{\bf x}_3-{\bf x}_2|^2-\frac{\beta^2}{24}|{\bf x}_3-{\bf x}_2|^3+\ldots\right],
\end{equation}
where an expansion of the exponent $e^{-\beta|{\bf x}_2-{\bf x}_3|}$ over $\beta|{\bf x}_2-{\bf x}_3|$ is used. It is equivalent to an expansion in
powers of $\sqrt{M_e/M_\mu}$. Whereas the first term $\beta^{-1}$ does not contribute, the second term in \eqref{eq:19} yields $-\nu_F\frac{35M_e}{24M_\mu}$.
In addition the third term in \eqref{eq:21} leads to the following integral:
\begin{equation}
\label{eq:22}
\int \psi^\ast_{\mu 0}({\bf x}_3)\sum_{n}\sqrt{2M_e(E_{\mu n}-E_{\mu 0})}\psi_{\mu n}({\bf x}_3)\psi^\ast_{\mu n}({\bf x}_2)({\bf x}_2\cdot{\bf x}_3)
\psi_{\mu 0}({\bf x}_2)d{\bf x}_2d{\bf x}_3=\frac{1}{3\alpha M_e}\left(\frac{M_e}{M_\mu}\right)^{3/2}S_{1/2},
\end{equation}
where we define
\begin{equation}
\label{eq:23}
S_{1/2}=\sum_{n}\left(\frac{E_{\mu n}-E_{\mu 0}}{R_\mu}\right)^{1/2}|\langle\mu 0|\frac{\bf x}{a_\mu}|\mu n\rangle|^2.
\end{equation}

Discrete and continuum states contributions to~\eqref{eq:23} are equal correspondingly \cite{HBES,VAF}:
\begin{equation}
\label{eq:23a}
S_{1/2}^{d}=\sum_{n}\frac{2^{8}n^6(n-1)^{2n-\frac{9}{2}}}{(n+1)^{2n+\frac{9}{2}}}=1.90695...,
\end{equation}
\begin{displaymath}
S_{1/2}^{c}=\int_0^\infty \frac{2^8kdk}{(k^2+1)^{9/2}(1-e^{-\frac{2\pi}{k}})}\left|\left(\frac{1+ik}{1-ik}\right)^{i/k}\right|=1.03111...  ,
\end{displaymath}
where $R_\mu=\frac{9}{2}M_\mu\alpha^2$.
Summing corrections in the first and second order PT we obtain total recoil correction to the coefficient $b$ in order $\alpha^4$:
\begin{equation}
\label{eq:24}
b_{rec}=\nu_F(1+\kappa_\mu)\left[-3\frac{M_e}{M_\mu}-\frac{8}{9}\frac{M_e^2}{M_\mu^2}\ln\frac{M_e}{M_\mu}+\frac{4}{9}\left(\frac{M_e}{M_\mu}\right)^{3/2}S_{1/2}+
\frac{8}{9}\frac{M_e^2}{M_\mu^2}\left(\frac{185}{64}-2\ln 2+\ln 3\right)\right].
\end{equation}

There are similar contributions to the coefficient $c$ in second order PT. In order to obtain it we have to use $\Delta H^{hfs}_0({\bf x}_e)=\frac{2\pi\alpha}{3}
\frac{g_eg_N}{m_em_p}\delta({\bf x}_e)$ in general expression \eqref{eq:12}. After evident simplifications recoil correction to $c$ can be written as
\begin{equation}
\label{eq:25}
c_1=\frac{4\pi\alpha}{3}\frac{g_eg_\mu}{m_em_p}\int \psi^\ast_{e0}(0)\tilde G_e(0,{\bf x}_1)V_\mu({\bf x}_1)\psi_{e0}({\bf x}_1)d{\bf x}_1.
\end{equation}
Appearing here the electron reduced Green's function with one zero argument has a form:
\begin{equation}
\label{eq:26}
\tilde G_e(0,{\bf x})=\sum_{n\not =0}^\infty\frac{\psi_{en}(0)\psi^\ast_{en}({\bf x})}{E_{e0}-E_{en}}=-\frac{2\alpha M_e^2}{\pi}e^{-2\alpha M_e x}
\left[\frac{1}{4\alpha M_e x}-\ln 4\alpha M_e x+\frac{5}{2}-C-2\alpha M_e x\right].
\end{equation}
The result of analytical integration is presented as an expansion in $M_e/M_\mu$:
\begin{equation}
\label{eq:27}
c_1=c_0\left[\frac{M_e}{M_\mu}+\frac{8}{9}\frac{M_e^2}{M_\mu^2}\left(\frac{1}{4}+\ln\frac{3}{2}-\ln\frac{M_e}{M_\mu}\right)\right]=
\Biggl\{{{8.467~MHz}\atop{22.302~MHz}}.
\end{equation}

\section{Effects of the vacuum polarization}

The vacuum polarization (VP) effects lead to the appearance of new terms in the
Hamiltonian which we denote $\Delta H_{VP}$ in \eqref{eq:1}. The ratio of the electron Compton wave length to the Bohr radius
in the subsystem $(\mu ^{6,7}_3Li)^{++}$: $Zm_\mu\alpha/m_e$=
$2.96185\ldots$ is not small value. So, we can not use for the calculation of VP effects an expansion over $\alpha$.
In this section we present a calculation of vacuum polarization corrections to hyperfine
structure in the first and second orders of perturbation theory. A modification of the Coulomb potentials
due to VP effects is described by the following relations \cite{t4,EGS}:
\begin{equation}
\label{eq:28}
\Delta V_{VP}^{eN}(x_e)=\frac{\alpha}{3\pi}\int_1^\infty
\rho(\xi)\left(-\frac{3\alpha}{x_e}\right) e^{-2m_e\xi
x_e}d\xi,~~~\rho(\xi)=\frac{\sqrt{\xi^2-1}(2\xi^2+1)}{\xi^4},
\end{equation}
\begin{equation}
\label{eq:29}
\Delta V_{VP}^{\mu N}(x_\mu)=\frac{\alpha}{3\pi}\int_1^\infty
\rho(\xi)\left(-\frac{3\alpha}{x_\mu}\right) e^{-2m_e\xi x_\mu}d\xi,
\end{equation}
\begin{equation}
\label{eq:30}
\Delta V_{VP}^{e\mu}(|{\bf x}_e-{\bf
x}_\mu|)=\frac{\alpha}{3\pi}\int_1^\infty
\rho(\xi)\frac{\alpha}{x_{e\mu}} e^{-2m_e\xi x_{e\mu}}d\xi,
\end{equation}
where $x_{e\mu}=|{\bf x}_e-{\bf x}_\mu|$. They give contributions to
hyperfine splitting in the second order perturbation theory and
are discussed below. In the first order perturbation theory the
contribution of the vacuum polarization is connected with a
modification of hyperfine splitting part of the Hamiltonian \eqref{eq:4}
(the amplitude in Fig.~\ref{fig:pic1}(a)). In the coordinate representation it is
determined by the integral expressions \cite{M1}:
\begin{equation}
\label{eq:31}
\Delta V_{VP}^{hfs,e\mu}({\bf x}_{e\mu})=-\frac{8\alpha}{3m_em_\mu}
({\bf S}_e\cdot {\bf S}_\mu)\frac{\alpha}{3\pi}\int_1^\infty
\rho(\xi)d\xi\left[\pi\delta({\bf
x_{e\mu}})-\frac{m_e^2\xi^2}{x_{e\mu}}e^{-2m_e\xi x_{e\mu}} \right],
\end{equation}
\begin{equation}
\label{eq:32}
\Delta V_{VP}^{hfs,eN}({\bf x}_e)=\frac{8\alpha g_N}{6m_em_p} ({\bf
S}_e\cdot {\bf I})\frac{\alpha}{3\pi}\int_1^\infty
\rho(\xi)d\xi\left[\pi\delta({\bf
x_e})-\frac{m_e^2\xi^2}{x_e}e^{-2m_e\xi x_e} \right].
\end{equation}

Averaging the potential \eqref{eq:31} over the wave function \eqref{eq:3} we obtain
the following contribution to the coefficient $b$:
\begin{equation}
\label{eq:33}
b_{vp}=\frac{8\alpha^2}{9m_em_\mu}\frac{(2\alpha M_e)^3(3\alpha
M_\mu)^3} {\pi^3}\int_1^\infty\rho(\xi)d\xi\int d{\bf x}_e\int d{\bf
x}_\mu e^{-6\alpha M_\mu x_\mu}e^{-4\alpha M_ex_e}\times
\end{equation}
\begin{displaymath}
\times\left[\pi\delta({\bf x_\mu}-{\bf x}_e)-\frac{m_e^2\xi^2}{|{\bf x}_\mu-{\bf x}_e|}
e^{-2m_e\xi|{\bf x}_\mu-{\bf x}_e|}\right].
\end{displaymath}
There are two integrals over the muon and electron coordinates in
\eqref{eq:33} which can be calculated analytically:
\begin{equation}
\label{eq:34}
I_1=\int d{\bf x}_e\int d{\bf x}_\mu e^{-6\alpha M_\mu x_\mu}e^{-4\alpha M_ex_e}\pi
\delta({\bf x_\mu}-{\bf x}_e)
=\frac{\pi^2}{(3\alpha M_\mu)^3\left(1+\frac{2M_e}{3M_\mu}\right)^3},
\end{equation}
\begin{equation}
\label{eq:35}
I_2=\int d{\bf x}_e\int d{\bf x}_\mu e^{-6\alpha M_\mu x_\mu}e^{-4\alpha M_ex_e}\frac{m_e^2\xi^2}
{|{\bf x}_\mu-{\bf x}_e|}e^{-2m_e\xi|{\bf x}_\mu-{\bf x}_e|}=
\end{equation}
\begin{displaymath}
=\frac{\pi^2m_e^2\xi^2}{(3\alpha M_\mu)^5}\frac{\left[\frac{4M_e^2}{9M_\mu^2}+\left(1+
\frac{m_e\xi}{3M_\mu\alpha}\right)^2+\frac{M_e}{3M_\mu}\left(6+\frac{4m_e\xi}{3M_\mu
\alpha}\right)\right]}{\left(1+\frac{2M_e}{3M_\mu}\right)^3\left(1+\frac{m_e\xi}
{3M_\mu\alpha}\right)^2\left(\frac{2M_e}{3M_\mu}+\frac{m_e\xi}{3M_\mu\alpha}\right)^2}.
\end{displaymath}
They are divergent separately in the subsequent integration over the
parameter $\xi$. But their sum is finite and can be written in the
integral form:
\begin{equation}
\label{eq:36}
b_{vp}=\nu_F\frac{2\alpha M_e}{9\pi
M_\mu\left(1+\frac{2M_e}{3M_\mu}\right)^3}\int_1^\infty\rho(\xi)d\xi
\frac{\left[\frac{2M_e}{3M_\mu}+2\frac{m_e\xi}{3M_\mu\alpha}\frac{2M_e}{3M_\mu}+
\frac{m_e\xi}{3M_\mu\alpha}\left(2+\frac{m_e\xi}{3M_\mu
\alpha}\right)\right]}{\left(1+\frac{m_e\xi}
{3M_\mu\alpha}\right)^2\left(\frac{2M_e}{3M_\mu}+\frac{m_e\xi}{3M_\mu\alpha}\right)^2}=\Biggl\{{{0.701~MHz}\atop{0.706~MHz}},
\end{equation}
The order of this contribution is determined by two small parameters
$\alpha$ and $M_e/M_\mu$ which are written explicitly. The
correction $b_{vp}$ is of the fifth order in $\alpha$ and the first
order in $M_e/M_\mu$. The contribution
of the muon vacuum polarization to hyperfine splitting is
negligibly small. One should expect that
two-loop vacuum polarization contributions to the hyperfine
structure are suppressed relative to the one-loop VP contribution by
the factor $\alpha/\pi$. This means that at present level of
accuracy we can neglect these corrections because their numerical
value is small. Higher orders of the perturbation
theory which contain one-loop vacuum polarization and the Coulomb
interaction \eqref{eq:2} lead to the recoil corrections of order
$\nu_F\alpha\frac{M_e^2}{M_\mu^2}\ln\frac{M_\mu}{M_e}$. Such terms
are included in the theoretical error.

Similar contribution to the coefficient $c$ of order $\alpha^6$ can
be found analytically using the potential \eqref{eq:32} ($\alpha_1=2\alpha M_e/m_e$):
\begin{equation}
\label{eq:37}
c_{vp}=\nu_F\frac{\alpha g_N m_\mu}{6\pi m_p}\frac{\sqrt{1-\alpha_1^2}(6\alpha_1+\alpha_1^3
-3\pi)+(6-3\alpha_1^2+6\alpha_1^4)\arccos\alpha_1}{3\alpha_1^3\sqrt{1-\alpha_1^2}}=\Biggl\{{{0.066~MHz}\atop{0.175~MHz}},
\end{equation}

Let us consider corrections of the electron vacuum polarization
\eqref{eq:28}-\eqref{eq:30} in the second order perturbation theory (SOPT) (the
amplitude in Fig.~\ref{fig:pic1}(b)). The contribution of the Coulomb
electron-nucleus interaction \eqref{eq:28} to the hyperfine splitting can be
written as follows:
\begin{equation}
\label{eq:38}
b_{vp,~SOPT}^{e-N}=\frac{4\pi\alpha g_eg_\mu}{3m_em_\mu}\int d{\bf
x}_1\int d{\bf x}_2\int d{\bf x}_3
\frac{\alpha}{3\pi}\int_1^\infty\rho(\xi)d\xi\psi^\ast_{\mu 0}({\bf
x}_3)\psi^\ast_{e 0}({\bf x}_3)\times
\end{equation}
\begin{displaymath}
\times\sum_{n,n'\not =0}^\infty\frac{\psi_{\mu n}({\bf x}_3) \psi_{e
n'}({\bf x}_3)\psi^\ast_{\mu n}({\bf x}_2)\psi^\ast_{e n'}({\bf
x}_1)}{E_{\mu 0}+E_{e0}-E_{\mu n}-E_{en'}}\left(\frac{-3\alpha}{x_1}\right)e^{-2m_e\xi x_1}\psi_{\mu
0}({\bf x}_2) \psi_{e0}({\bf x}_1),
\end{displaymath}
where the indices at the coefficient $b$ indicate vacuum
polarization contribution (VP) in the second order PT (SOPT) when
the electron-nucleus Coulomb VP potential is considered. The
summation in \eqref{eq:38} is carried out over the complete system of the
eigenstates of the electron and muon excluding the state with
$n,n'=0$. The computation of the expression \eqref{eq:38} is simplified with
the use of the orthogonality condition for the muon wave functions:
\begin{equation}
\label{eq:39}
b_{vp,~SOPT}^{e-N}=\nu_F\frac{2\alpha M_e^2}{9\pi
M_\mu^2}\int_1^\infty \rho(\xi)d\xi\int _0^\infty x_3^2 dx_3\int
_0^\infty x_1 dx_1
e^{-x_1\frac{2M_e}{3M_\mu}\left(1+\frac{m_e\xi}{2\alpha M_e}\right)}
e^{-x_3\left(1+\frac{2M_e}{3M_\mu}\right)}\times
\end{equation}
\begin{displaymath}
\Bigl[\frac{3M_\mu}{2M_ex_>}-\ln\left(\frac{2M_e}{3M_\mu}x_<\right)-\ln
\left(\frac{2M_e}{3M_\mu}x_>\right)+Ei\left(\frac{2M_e}{3M_\mu}x_<\right)+
\frac{7}{2}-2C-\frac{M_e}{3M_\mu}(x_1+x_3)+
\end{displaymath}
\begin{displaymath}
+\frac{1-e^{\frac{2M_e}{3M_\mu}x_<}}{\frac{2M_e}{3M_\mu}x_<}\Bigr]=\Biggl\{{{1.136~MHz}\atop{1.137~MHz}},
\end{displaymath}
It is necessary to emphasize that the transformation of
the expression \eqref{eq:38} into \eqref{eq:39} is carried out with the use of \eqref{eq:16}.

The contribution \eqref{eq:39} has the same order of the magnitude
$O(\alpha^5\frac{M_e}{M_\mu})$ as previous correction \eqref{eq:36} in
the first order perturbation theory. Similar calculation can be
performed in the case of muon-nucleus Coulomb vacuum polarization
potential \eqref{eq:29}. An intermediate electron state is the $1S$-state and
the reduced Coulomb Green's function of the system transforms to the Green's function of the muon.
The correction of the operator \eqref{eq:29} to the hyperfine splitting
(the coefficient $b$) is obtained in the following integral form:
\begin{equation}
\label{eq:40}
b_{vp~SOPT}^{\mu-N}
=\nu_F\frac{\alpha}{3\pi}\int_1^\infty\rho(\xi)d\xi\int_0^\infty
x_3^2dx_3\int_0^\infty x_2dx_2
e^{-x_3\left(1+\frac{2M_e}{3M_\mu}\right)}e^{-x_2\left(1+\frac{m_e\xi}
{3M_\mu\alpha}\right)}\times
\end{equation}
\begin{displaymath}
\times\left[\frac{1}{x_>}-\ln x_>-\ln x_<+Ei (x_<)+\frac{7}{2}-2C
-\frac{x_2+x_3}{2}+\frac{1-e^{x_<}}{x_<}\right]=\Biggl\{{{0.694~MHz}\atop{0.693~MHz}},
\end{displaymath}
The vacuum polarization correction to HFS which is determined by the
operator \eqref{eq:30} in the second order perturbation theory is the most
difficult for the calculation. Indeed, in this case we have to
consider the intermediate excited states both for the muon and
electron. We have divided total contribution into two parts. The first part in which the
intermediate muon is in the $1S$-state can be written as:
\begin{equation}
\label{eq:41}
b_{vp,~SOPT}^{\mu-e}(n=0)=\frac{256\alpha^2(2\alpha
M_e)^3(3\alpha M_\mu)^3}{9m_em_\mu} \int_0^\infty x_3^2dx_3e^{-\alpha(2M_e+6M_\mu)x_3}\times
\end{equation}
\begin{displaymath}
\times\int_0^\infty x_1^2 dx_1 e^{-2\alpha M_ex_1}\int_1^\infty\rho(\xi)
d\xi\Delta V_{VP~\mu}(x_1)G_e(x_1,x_3),
\end{displaymath}
where the function $V_{VP~\mu}(x_1)$ is equal
\begin{equation}
\label{eq:42}
\Delta V_{VP~\mu}(x_1)=\int d{\bf x}_2 e^{-6\alpha M_\mu
x_2}\frac{(3\alpha M_\mu)^3} {\pi}\frac{\alpha}{|{\bf x}_1-{\bf
x}_2|}e^{-2m_e\xi|{\bf x}_1-{\bf x}_2|}=
\end{equation}
\begin{displaymath}
=\frac{108\alpha^4 M^3_\mu}{x_1(36\alpha^2 M_\mu^2-4m_e^2\xi^2)^2}
\left[12\alpha M_\mu\left(e^{-2m_e\xi x_1}-e^{-6\alpha M_\mu
x_1}\right)+x_1(4m_e^2\xi^2-36\alpha^2 M_\mu^2)e^{-6\alpha M_\mu
x_1}\right].
\end{displaymath}
After substitution \eqref{eq:42} into \eqref{eq:41} numerical integration gives
the result:
\begin{equation}
\label{eq:43}
b_{vp,~SOPT}^{\mu-e}(n=0)=\Biggl\{{{-0.310~MHz}\atop{-0.309~MHz}},
\end{equation}
Second part of the vacuum polarization correction to the hyperfine
splitting due to the electron-muon interaction can be presented as follows:
\begin{equation}
\label{eq:44}
b_{vp,~SOPT}^{\mu-e}(n\not=0)=-\frac{4\alpha^2}{9}\frac{g_eg_\mu}{m_em_\mu}\int d{\bf x}_3\int d{\bf
x}_2\int_1^\infty\rho(\xi)d\xi \psi^\ast_{\mu 0}({\bf x}_3)\psi^\ast_{e 0}({\bf x}_3)\times
\end{equation}
\begin{displaymath}
\times\sum_{n\not =0}\psi_{\mu n}({\bf x}_3)\psi^\ast_{\mu n}({\bf
x}_2) \frac{M_e}{2\pi}\frac{e^{-{\beta}|{\bf x}_3-{\bf
x}_1|}}{|{\bf x}_3-{\bf x}_1|}\frac{\alpha}{|{\bf x}_2-{\bf
x}_1|}e^{-2m_e\xi|{\bf x}_2-{\bf x}_1|} \psi_{\mu 0}({\bf
x}_2)\psi_{e 0}({\bf x}_1),
\end{displaymath}
where we have replaced exact electron Coulomb Green's function by free electron Green's function.
We also replace the electron wave functions by their values at the origin neglecting higher
order recoil corrections. After that the integration over ${\bf x}_1$ can be done analytically:
\begin{equation}
\label{eq:45}
J=\int d{\bf x}_1\frac{e^{-{\beta}|{\bf x}_3-{\bf x}_1|}}{|{\bf
x}_3-{\bf x}_1|} \frac{e^{-2m_e\xi|{\bf x}_2-{\bf x}_1|}}{|{\bf
x}_2-{\bf x}_1|}= -\frac{4\pi}{|{\bf x}_3-{\bf x}_2|}\frac{1}{{\beta}^2-4m_e^2\xi^2}\left[e^{-{\beta}|{\bf x}_3-{\bf
x}_2|}-e^{-2m_e\xi|{\bf x}_3-{\bf x}_2|}\right]=
\end{equation}
\begin{displaymath}
=2\pi\Biggl[\frac{\left(1-e^{-2m_e\xi|{\bf x}_3-{\bf x}_2|}\right)}
{2m_e^2\xi^2|{\bf x}_3-{\bf x}_2|}- \frac{{\beta}}{2m_e^2\xi^2}+\frac{\left(1-e^{-2m_e\xi|{\bf x}_3-{\bf
x}_2|}\right){\beta}^2} {8m_e^4\xi^4|{\bf x}_3-{\bf x}_2|}+\frac{{\beta}^2|{\bf x}_3-{\bf x}_2|}{4m_e^2\xi^2}-
\end{displaymath}
\begin{displaymath}
-\frac{{\beta}^3}{8m_e^4\xi^4}-\frac{{\beta}^3({\bf x}_3-{\bf x}_1)^2}{12m_e^2\xi^2}+...\Biggr],
\end{displaymath}
where we perform an expansion of the first exponent in square brackets in powers of ${\beta}|{\bf x}_3-{\bf x}_2|$.
For further transformation the completeness condition is useful:
\begin{equation}
\label{eq:46}
\sum_{n\not =0}\psi_{\mu n}({\bf x}_3)\psi_{\mu n}^\ast({\bf x}_2)=
\delta({\bf x}_3-{\bf x}_2)-\psi_{\mu 0}({\bf x}_3)\psi_{\mu
0}^\ast({\bf x}_2).
\end{equation}
The wave function orthogonality leads to the zero results for the
second and fifth terms in the square brackets of \eqref{eq:45}. The first
term in \eqref{eq:45} gives the leading order contribution in two small
parameters $\alpha$ and $M_e/M_\mu$ ($\gamma=m_e\xi/3\alpha M_\mu$):
\begin{equation}
\label{eq:47}
b_{vp,~SOPT}^{ \mu-e}(n\not=0)=b_{11}+b_{12}=\Biggl\{{{-0.432~MHz}\atop{-0.431~MHz}},~
b_{11}=-\frac{3\alpha^2M_e}{8m_e}\nu_F,
\end{equation}
\begin{equation}
\label{eq:48}
b_{12}=\nu_F\frac{\alpha^2M_e}{24\pi m_e}\int_1^\infty\frac{\rho(\xi)d\xi}{\xi}
\frac{[16+\gamma(5\gamma(\gamma+4)+29)]}{(1+\gamma)^4}.
\end{equation}
Numerical value of the sum $b_{11}+b_{12}$ is included in Table~\ref{tb1}. It is important to calculate also the
contributions of other terms of the expression \eqref{eq:45} to the hyperfine
splitting. Taking the fourth term in \eqref{eq:45} which is proportional to ${\beta}^2=2M_e(E_{\mu n}-E_{\mu 0})$
we have performed the sequence of transformations in the coordinate representation:
\begin{equation}
\label{eq:49}
\sum_{n=0}^\infty E_{\mu n}\int d{\bf x}_2\int d{\bf x}_3\psi_{\mu
0}^\ast({\bf x}_2) \psi_{\mu n}({\bf x}_3)\psi_{\mu n}^\ast({\bf
x}_2)|{\bf x}_3-{\bf x}_2|\psi_{\mu 0}({\bf x}_2)=
\end{equation}
\begin{displaymath}
=\int d{\bf x}_2\int d{\bf x}_3\delta({\bf x}_3-{\bf x}_2)
\left[-\frac{\nabla^2_3}{2M_\mu}|{\bf x}_3-{\bf x}_2|\psi_{\mu 0}^\ast({\bf x}_3)\right]
\psi_{\mu 0}({\bf x}_2).
\end{displaymath}
Evidently, we have the divergent expression in \eqref{eq:49} due to the
presence of the $\delta$-function. The same divergence occurs in the
other term containing ${\beta}^2$ in the square brackets of
\eqref{eq:45}. But their sum is finite and can be calculated analytically with the result:
\begin{equation}
\label{eq:50}
b_{{\beta}^2}=\nu_F\frac{3\alpha^2
M_e^2}{8m_eM_\mu}\left(1+\frac{5}{8}\frac{\alpha^2 M_\mu^2}{m_e^2}\right).
\end{equation}
Numerical value of this correction is essentially smaller than the
leading order term. Other terms in \eqref{eq:45} give negligibly small corrections.

\begin{figure}[htbp]
\centering
\includegraphics{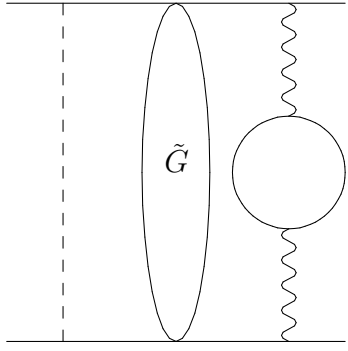}
\caption{Vacuum polarization effects in the second order
perturbation theory. The dashed line represents the first part of
the potential $\Delta H$ (3). The wave line represents the hyperfine
part of the Breit potential.}
\label{fig:pic2}
\end{figure}

The Coulomb vacuum polarization \eqref{eq:28} does not contain the muon coordinate, so, its contribution to
the coefficient $c$ in the second order PT can be derived taking $n=0$ for the muon state in
the Coulomb Green's function. Moreover, the $\delta({\bf x}_e)$ function in \eqref{eq:4} leads to the
appearance of the electron Green's function with one zero argument. Corresponding value of hyperfine splitting is equal
\begin{equation}
\label{eq:51}
c_{vp,~SOPT}^{e-N}=\nu_F\frac{\alpha m_\mu g_eg_N}{4\pi m_p}\int_1^\infty\rho(\xi)d\xi
\frac{2a_1^2+3a_1+2a_1\ln a_1-2}{2a_1^3}=\Biggl\{{{0.104~MHz}\atop{0.274~MHz}}.
\end{equation}
The vacuum polarization in the Coulomb $(\mu-N)$ interaction does
not contribute to $c$ in SOPT because of the orthogonality of the
muon wave functions. Let us consider correction to the coefficient $c$
which arises from \eqref{eq:30} in SOPT. Only intermediate muon state with $n=0$ in
the Green's function gives the contribution in this case. Using \eqref{eq:26}
we make integration over electron coordinates and present this
correction in the form ($\gamma=m_e\xi/3\alpha M_\mu$, $\gamma_1=2M_e/3M_\mu$):
\begin{equation}
\label{eq:52}
c_{vp,~SOPT}^{e-\mu}=-\nu_F\frac{2\alpha m_\mu g_NM_e^2}{27\pi
m_pM_\mu^2}\int_1^\infty\frac{\rho(\xi)d\xi}{(1-\gamma^2)^2}\int_0^\infty xe^{-\gamma_1x}dx
\times
\end{equation}
\begin{displaymath}
\times\left[e^{-\gamma x}-e^{-x}+\frac{x}{2}e^{-x}\left(\gamma^2-1\right)\right]
\left[\frac{1}{\gamma_1 x}-\ln\gamma_1 x+\frac{5}{2}-C-\frac{1}{2}\gamma_1 x\right]=\Biggl\{{{-0.018~MHz'}\atop{-0.047~MHz}},
\end{displaymath}

There exist another contributions of the second order perturbation
theory in which we have the vacuum polarization perturbation
connected with hyperfine splitting parts of the Breit potential
\eqref{eq:31}-\eqref{eq:32} (see Fig.~\ref{fig:pic2}). Other perturbation potential in this case is
determined by the first term of relation \eqref{eq:2}. We can divide the HFS
correction of \eqref{eq:32} into two parts. One part with $n=0$ corresponds
to muon ground state. The other part with $n\not =0$ accounts for
excited muon states. The $\delta$-function term in \eqref{eq:31} gives
the following contribution to HFS at $n=0$:
\begin{equation}
\label{eq:53}
b_{vp,~SOPT}^{(11)}(n=0)=\nu_F\frac{\alpha}{3\pi}\int_1^\infty\rho(\xi)d\xi
\frac{11M_e}{24 M_\mu}.
\end{equation}
Obviously, this integral in the variable $\xi$ is divergent. So, we
have to consider the contribution of the second term of the
potential \eqref{eq:31} to hyperfine splitting which is determined by
more complicated expression:
\begin{equation}
\label{eq:54}
b_{vp,~SOPT}^{(12)}(n=0)=\frac{16\alpha^2m_e^2}{9\pi
m_em_\mu}\int_1^\infty \rho(\xi)\xi^2d\xi\int d{\bf x}_3\psi_{e
0}({\bf x}_3)\Delta V_1({\bf x}_3)\times
\end{equation}
\begin{displaymath}
\times\int \sum_{n'\not =0}
\frac{\psi_{e n'}({\bf x}_3)\psi^\ast_{e n'}({\bf x}_1)}{E_{e0}-E_{en'}}\Delta V_2({\bf x}_1)
\psi_{e 0}({\bf x}_1)d{\bf x}_1,
\end{displaymath}
where $\Delta V_1({\bf x}_3)$ is defined in \eqref{eq:42} and $\Delta V_2({\bf x}_1)$ in \eqref{eq:15}.
Integrating in \eqref{eq:54} over all coordinates we obtain
the following result in the leading order in the ratio $(M_e/M_\mu)$:
\begin{equation}
\label{eq:57}
b_{vp,~SOPT}^{(12)}(n=0)=-\nu_F\frac{m_e}{M_e}\frac{M_e^2}{216\pi
M_\mu^2}\int_1^\infty\rho(\xi)\xi
d\xi\frac{32+63\gamma+44\gamma^2+11\gamma^3}{(1+\gamma)^4}.
\end{equation}
This integral also has the divergence at large values of the
parameter $\xi$. But the sum of integrals \eqref{eq:53} and \eqref{eq:57} is finite:
\begin{equation}
\label{eq:58}
b_{vp,~SOPT}^{(11)}(n=0)+b_{vp,~SOPT}^{(12)}(n=0)=\nu_F\frac{\alpha M_e}{72\pi
M_\mu}\int_1^\infty\rho(\xi)d\xi\frac{11+12\gamma+3\gamma^2}{(1+\gamma)^4}=\Biggl\{{{0.067~MHz}\atop{0.067~MHz}}.
\end{equation}

Let us consider now the terms in the coefficient $b$ with $n\not
=0$. The delta-like term of the potential \eqref{eq:32} gives the
following contribution to the HFS:
\begin{equation}
\label{eq:59}
b_{vp,~SOPT}^{(21)}(n\not =0)=\nu_F\frac{\alpha}{3\pi}\int_1^\infty\rho(\xi)d\xi\left(-\frac{35M_e}{24M_\mu}\right).
\end{equation}
Another correction from the second term of the expression \eqref{eq:32} can
be simplified after the replacement of exact electron Green's
function by free electron Green's function:
\begin{equation}
\label{eq:60}
b_{vp,~SOPT}^{(22)}(n\not=0)=-\frac{16\alpha^3M_em_e^2}{9\pi m_em_\mu}
\int_1^\infty\rho(\xi)\xi^2d\xi\int d{\bf x}_2\int d{\bf x}_3\times
\end{equation}
\begin{displaymath}
\times\int d{\bf x}_4\psi_{\mu 0}^\ast({\bf x}_4)
\frac{e^{-2m_e\xi|{\bf x}_3-{\bf x}_4|}}{|{\bf x}_3-{\bf x}_4|}\sum_{n\not=0}^\infty
\psi_{\mu n}({\bf x}_4)\psi_{\mu n}({\bf x}_2)|{\bf x}_3-{\bf x}_2|\psi_{\mu 0}({\bf x}_2)
\end{displaymath}
Analytical integration in \eqref{eq:60} over all coordinates leads to
the result:
\begin{equation}
\label{eq:61}
b_{vp,~SOPT}^{(22)}(n\not=0)=-\nu_F\frac{2\alpha M_e}{9\pi
M_\mu}\int_1^\infty\rho(\xi)d\xi[
\frac{1}{\gamma}-\frac{1}{(1+\gamma)^4}(4+\frac{1}{\gamma}+10\gamma+\frac{215\gamma^2}{16}+
\frac{35\gamma^3}{4}+\frac{35\gamma^4}{16})].
\end{equation}
The sum of expressions \eqref{eq:59} and \eqref{eq:61} gives again the finite
contribution to the hyperfine splitting:
\begin{equation}
\label{eq:62}
b_{vp,~SOPT}^{(21)}(n\not=0)+b_{vp,~SOPT}^{(22)}(n\not=0)=
\end{equation}
\begin{displaymath}
=-\nu_F\frac{2\alpha M_e}{9\pi M_\mu}
\int_1^\infty\rho(\xi)d\xi\frac{35+76\gamma+59\gamma^2+16\gamma^3}{16(1+\gamma)^4}=\Biggl\{{{-0.432~MHz}\atop{-0.431~MHz}}.
\end{displaymath}
Absolute values of calculated VP corrections \eqref{eq:37}, \eqref{eq:41}, \eqref{eq:43},
\eqref{eq:44}, \eqref{eq:46}, \eqref{eq:58}, \eqref{eq:62} are sufficiently large ,
their summary contribution to the hyperfine splitting (see Table~\ref{tb1}) is small because they have different signs.

The hyperfine splitting interaction \eqref{eq:32} gives the contributions to
the coefficient $c$ in second order PT. Since the muon coordinate
does not enter into the expression \eqref{eq:32}, we should set $n=0$ for the
muon intermediate states in the Green's function. The initial formula for this correction is
\begin{equation}
\label{eq:63}
c_{vp,~SOPT}=\frac{8\alpha^3 g_N}{9\pi m_e
m_p}\int_1^\infty\rho(\xi)d\xi\int d{\bf x}_1\int d {\bf x}_3\int
d{\bf x}_4|\psi_{\mu 0}({\bf x}_3)|^2\psi^\ast_{e0}({\bf
x}_4)\psi_{e0}({\bf x}_1)\times
\end{equation}
\begin{displaymath}
\times\left[\frac{1}{|{\bf x}_3-{\bf
x}_4|}-\frac{1}{x_4}\right]G_e({\bf x}_4,{\bf
x}_1)\left(\pi\delta({\bf x}_1)-\frac{m_e^2\xi^2}{x_1}e^{-2m_e\xi
x_1}\right).
\end{displaymath}
The integration over ${\bf x}_3$ can be done analytically as in
\eqref{eq:15}. Then it is useful to divide \eqref{eq:63} into two parts. The
coordinate integration in the first term with the $\delta$ -
function is performed by means of \eqref{eq:27}. In the second term of \eqref{eq:63}
we use the electron Green's function in the form \eqref{eq:16}. The summary
result can be presented in the integral form in the leading order in $M_e/M_\mu$:
\begin{equation}
\label{eq:64}
c_{vp,~SOPT}=\nu_F\frac{\alpha g_N m_\mu M_e}{18\pi m_p
M_\mu}\int_1^\infty\rho(\xi)d\xi\frac{3+2\frac{m_e\xi}{3\alpha
M_\mu}}{(1+\frac{m_e\xi}{3\alpha M_\mu})^2}=\Biggl\{{{0.017~MHz}\atop{0.044~MHz}}.
\end{equation}

\section{Nuclear structure and recoil effects}

\begin{figure}[htbp]
\centering
\includegraphics{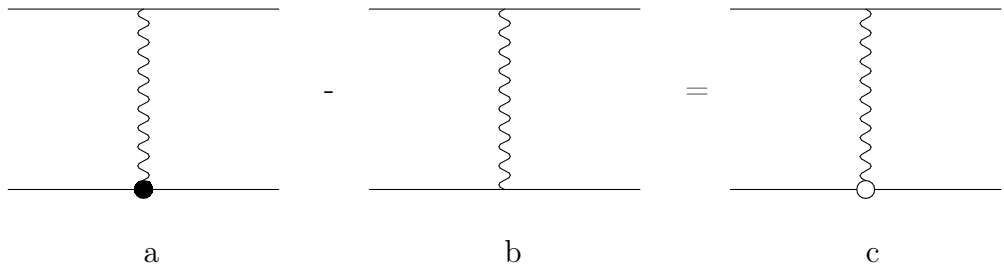}
\caption{Nuclear structure correction to coefficient $c$ in
$1\gamma$ interaction. The bold point represents the nuclear vertex
operator. The wave line represents the hyperfine part of the Breit
potential.}
\label{fig:pic3}
\end{figure}

Another set of significant corrections to hyperfine splitting of
muonic lithium ions which we study in this work is determined by the
nuclear structure and recoil \cite{GY,BY,M2004,M3,fmms}. We describe the charge and
magnetic moment distributions of $Li$ nucleus by means of two
form factors $G_E(k^2)$ and $G_M(k^2)$ for which we use the dipole
parameterization:
\begin{equation}
\label{eq:65}
G_E(k^2)=\frac{1}{\left(1+\frac{k^2}{\Lambda^2}\right)^2},~G_M(k^2)=\frac{G(0)}{\left(1+\frac{k^2}{\Lambda^2}\right)^2},~
G(0)=g_N\frac{m_N}{Zm_p}.
\end{equation}
where the parameter $\Lambda$ is related with the nucleus charge radius $r_N$: $\Lambda=\sqrt{12}/r_N$.
In $1\gamma$ - interaction the nuclear structure correction to the coefficient $c$ is determined by
the amplitudes shown in Fig.~\ref{fig:pic3}. Purely point contribution in Fig.~\ref{fig:pic3}(b)
leads to the HFS value \eqref{eq:11}. Then the nuclear structure correction is given by
\begin{equation}
\label{eq:66}
c_{str,~1\gamma}=\nu_F\frac{g_eg_Nm_\mu}{4m_p}\left[\int \frac{G_M(x)}{G_M(0)}e^{-4\alpha M_e x}
d{\bf x}-1\right]=\Biggl\{{{-0.283~MHz}\atop{-0.707~MHz}}.
\end{equation}

\begin{figure}[htbp]
\centering
\includegraphics{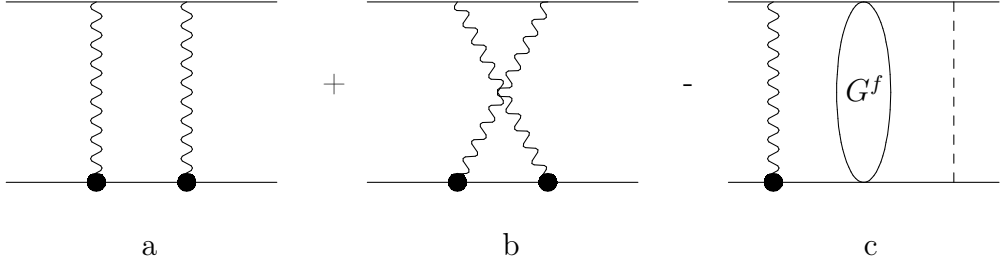}
\caption{Nuclear structure corrections to coefficient $c$ in
$2\gamma$ interactions. The bold point represents the nuclear vertex
operator. The wave line represents the hyperfine part of the Breit
potential. The dashed line corresponds to the Coulomb potential.}
\label{fig:pic4}
\end{figure}

\begin{figure}[htbp]
\centering
\includegraphics{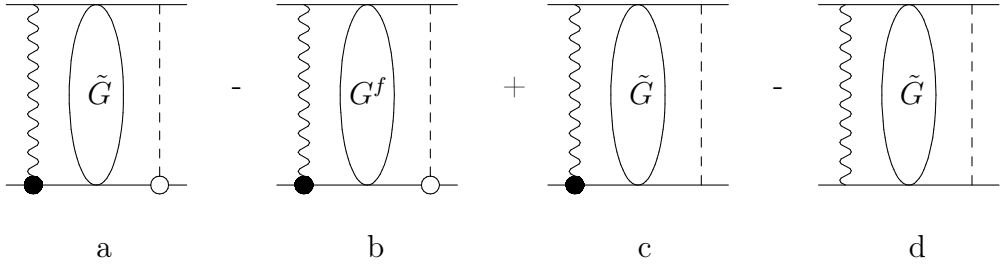}
\caption{Nuclear structure corrections to coefficient $c$ in second
order PT. The bold point represents the nuclear vertex operator. The
wave line represents the hyperfine part of the Breit potential. The dashed
line corresponds to the Coulomb potential. $\tilde G$ is the reduced
Coulomb Green's function.}
\label{fig:pic5}
\end{figure}

Two-photon $(e-N)$ interaction amplitudes (see Fig.~\ref{fig:pic4}) give the
contribution to HFS of order $\alpha^5$. It can be presented in
terms of the form factors $G_E$ and $G_M$ with the account of subtraction term as follows:
\cite{M3,M2000}:
\begin{equation}
\label{eq:67}
c_{str,~2\gamma}=\nu_F\frac{3\alpha m_\mu g_e g_N}{2\pi^2 m_p}\int \frac{d{\bf p}}{p^4}\frac{G_M(p)}{G_M(0)}\left[G_E(p)-1\right],
\end{equation}
where the subtraction term contains magnetic form factor $G_M(p)$. Using the dipole parameterization \eqref{eq:65} we
can present last integral in analytical form:
\begin{equation}
\label{eq:68}
c_{str,~2\gamma}=-\nu_F\frac{33\alpha M_em_\mu g_e g_N}{16m_p\Lambda}=\Biggl\{{{-0.195~MHz}\atop{-0.486~MHz}}.
\end{equation}
Other parts of the iteration contribution $\langle V_{1\gamma}\times G^f\times V_{1\gamma}\rangle_{str}^{hfs}$
are used in the second order PT (see Fig.~\ref{fig:pic5}).

\begin{figure}[htbp]
\centering
\includegraphics{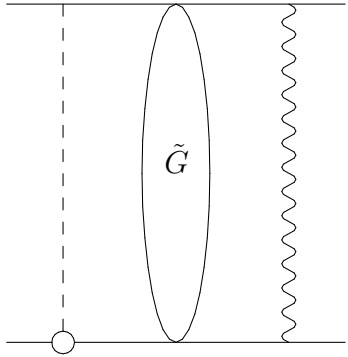}
\caption{Nuclear structure correction to coefficient $b$ in the
second order perturbation theory. The wave line represents the
hyperfine $(e-\mu)$ interaction. $\tilde G$ is the reduced Coulomb
Green's function.}
\label{fig:pic6}
\end{figure}

The nuclear structure corrections to the coefficient $c$ in second
order PT are presented in Fig.~\ref{fig:pic5}. Here we have two different
contributions. First contribution is related with amplitudes in
Fig.~\ref{fig:pic5}(a,b) when hyperfine part of one perturbation is
determined by magnetic form factor $G_M$ and the other perturbation
is described by the nucleus charge radius $r_N$:
\begin{equation}
\label{eq:70}
\Delta V^C_{str, e-N}({\bf r})=\frac{2}{3}\pi Z\alpha r^2_N\delta({\bf r}).
\end{equation}
The general integral structure of this correction and its numerical
value are as follows ($a_2=4\alpha M_e/\Lambda$):
\begin{equation}
\label{eq:71}
c_{1,str,~SOPT}^{e-N}=-\nu_F\frac{\alpha^2 r_N^2M_e^2g_eg_N m_\mu}{m_p}\int_0^\infty x^2 dx
e^{-x(1+a_2)}\left(-\ln a_2x+\frac{5}{2}-C-\frac{1}{2}a_2x\right)=
\end{equation}
\begin{displaymath}
=\Biggl\{{{-0.0003~MHz}\atop{-0.0008~MHz}},
\end{displaymath}
Numerical value of the contribution $c_{1,str,~SOPT}^{e-N}$ is
obtained by means of the charge radii of nucleus $^{6,7}_3Li$
$r(^6_3Li)=2.589(39)$ fm and $r(^7_3Li)=2.444(42)$ fm \cite{sick2011}. The second nuclear structure
contribution from amplitudes in Fig.~\ref{fig:pic5}(c,d) is evaluated by means
of the potential $\Delta H$ \eqref{eq:2} and the nucleus magnetic form factor.
For the amplitude in Fig.~\ref{fig:pic5}(c) we make the integration over the muon
coordinate in the muon state with $n=0$ and present the correction to
the coefficient $c$ in the form ($a_3=6\alpha M_\mu/\Lambda$):
\begin{equation}
\label{eq:72}
c_{2,str,~SOPT}^{e-N}+c_1=\nu_F\frac{2\alpha^2M_e^2m_\mu g_eg_N}{m_p\Lambda^2}\int_0^\infty x_1^2dx_1
e^{-x_1(1+a_2)}\int_0^\infty x_2dx_2\left(1+\frac{1}{2}a_3x_2\right)e^{-x_2(a_2+a_3)}\times
\end{equation}
\begin{displaymath}
\left[\frac{1}{a_2x_>}-\ln(a_2x_>)-\ln(a_2x_<)+Ei(a_2x_<)+\frac{7}{2}-2C-\frac{1}{2}a_2(x_1+x_2)+
\frac{1-e^{a_2x_<}}{a_2x_<}\right]=
\end{displaymath}
\begin{displaymath}
=\Biggl\{{{8.314~MHz}\atop{21.918~MHz}}.
\end{displaymath}
Subtracting the point contribution $c_1$ \eqref{eq:27} we find
\begin{equation}
\label{eq:72a}
c_{2,str,~SOPT}^{e-N}=\Bigl\{{{-0.153~MHz}\atop{-0.384~MHz}}.
\end{equation}

There is the nuclear structure contribution to the coefficient $b$
in second order PT which is presented in Fig.~\ref{fig:pic6}. If we consider the
Coulomb interaction between the muon and nucleus, then the structure
correction takes on the form:
\begin{equation}
\label{eq:73}
b_{str}^{\mu-N}=\frac{32\pi^2\alpha^2}{3m_em_\mu}r_N^2
\frac{1}{\sqrt{\pi}}\left(3\alpha M_\mu\right)^{3/2} \int d{\bf
x}_3\psi^\ast_{\mu 0}({\bf x}_3)|\psi_{e0}({\bf x}_3)|^2G_\mu({\bf
x}_3,0,E_{\mu 0}).
\end{equation}
After that an analytical integration over the coordinate ${\bf
x}_3$ in \eqref{eq:73} can be carried out using the representation of the
muon Green's function similar to expression \eqref{eq:26}. The result of the
integration of order $O(\alpha^6)$ is written as an expansion in the
ratio $M_e/M_\mu$:
\begin{equation}
\label{eq:74}
b_{str}^{\mu-N}=-\nu_F 24\alpha^2M_\mu^2
r_N^2\left(\frac{M_e}{M_\mu}-
\frac{22}{9}\frac{M_e^2}{M_\mu^2}+\ldots\right)=\Biggl\{{{-0.416~MHz}\atop{-0.372~MHz}}.
\end{equation}
The same approach can be used in the calculation of the structure correction to
electron-nucleus interaction. The electron feels as well the
distribution of the nucleus electric charge. The corresponding
contribution of the nuclear structure effect to hyperfine
splitting is determined by the expression:
\begin{equation}
\label{eq:75}
b_{str}^{e-N}=\frac{32\pi^2\alpha^2}{3m_em_\mu}r_N^2 \int d{\bf x}_1
\int d{\bf x}_3|\psi^\ast_{\mu 0}({\bf x}_3)|^2\psi_{e0}({\bf
x}_3)G_e({\bf x}_3,{\bf x}_1,E_{e 0}) \psi_{e0}({\bf
x}_1)\delta({\bf x}_1).
\end{equation}
Performing an analytical integration in \eqref{eq:75} we obtain the
following series:
\begin{equation}
\label{eq:76}
b_{str}^{e-N}=-\nu_F 6\alpha^2
M_eM_\mu r_N^2\left[1-\frac{4M_e}{3M_\mu}\ln\frac{2M_e}{3M_\mu}+
\frac{4M_e^2}{9M_\mu^2}\left(6\ln\frac{2M_e}{3M_\mu}-4\right)+\ldots\right]
=\Biggl\{{{-0.109~MHz}\atop{-0.098~MHz}}.
\end{equation}
We have included in Table~\ref{tb1} total nuclear structure contribution
to the coefficient $b$ which is equal to the sum of numerical values \eqref{eq:74} and \eqref{eq:76}.

\begin{figure}[htbp]
\centering
\includegraphics{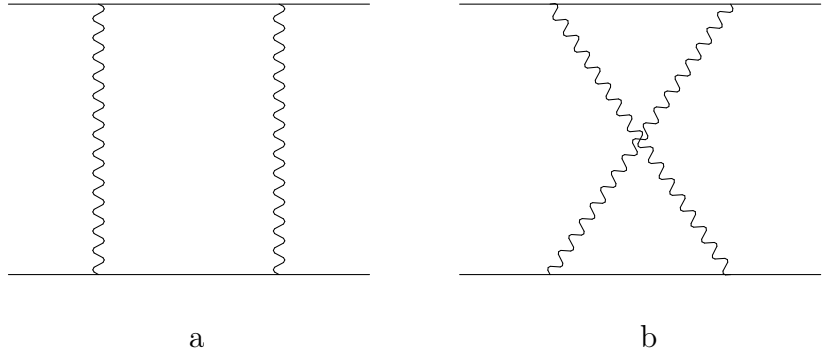}
\caption{Two photon exchange amplitudes in the electron-muon hyperfine
interaction.}
\label{fig:pic7}
\end{figure}

Special attention has to be given to the recoil corrections
connected with two-photon exchange diagrams shown in Fig.~\ref{fig:pic7} in the
case of the electron-muon interaction.  The leading order recoil
contribution to the interaction operator between the muon and
electron is determined by the expression \cite{EGS,Chen,RA}:
\begin{equation}
\label{eq:77}
\Delta V_{rec,\mu -e}^{hfs}({\bf x}_{\mu
e})=-8\frac{\alpha^2}{m_\mu^2-m_e^2}\ln\frac{m_\mu}{m_e}({\bf s}_\mu
{\bf s}_e) \delta({\bf x}_{\mu e}).
\end{equation}
Averaging the potential $\Delta V_{rec,\mu -e}^{hfs}$ over the wave
functions \eqref{eq:3} we obtain the recoil correction to the coefficient
$b$:
\begin{equation}
\label{eq:78}
b_{rec}^{\mu-e}=\nu_F\frac{3\alpha}{\pi}\frac{m_em_\mu}{m_\mu^2-m_e^2}\ln\frac{m_\mu}
{m_e}\frac{1}{\left(1+\frac{2M_e}{3M_\mu}\right)^3}=\Biggl\{{{6.430~MHz}\atop{6.431~MHz}}.
\end{equation}
There exist also two-photon interactions between the bound
particles of muonic lithium ions when one hyperfine photon transfers
the interaction from the electron to muon and another Coulomb photon
from the electron to the nucleus (or from the muon to the nucleus).
Supposing that these amplitudes give smaller contribution to
hyperfine splitting we included them in the theoretical error.

The first-order recoil correction $O(M_e/M_{Li})$ has a contribution from intermediate
states in which the muon and electron are excited to P-states:
\begin{equation}
\label{eq:78a}
\Delta b_{rec,SOPT}=-\frac{32\Pi\alpha^3 M_eM_\mu}{m_em_\mu M_{Li}}\int d{\bf x}_3\int d{\bf x}_2\int d{\bf x}_1
\Psi^\ast_{\mu 0}({\bf x}_3)\Psi^\ast_{e 0}({\bf x}_3)\times
\end{equation}
\begin{displaymath}
\times\sum_{n,n'\not=0}\frac{\Psi_{\mu n}({\bf x}_3)\Psi_{e n'}({\bf x}_3)\Psi^\ast_{\mu n}({\bf x}_2)\Psi^\ast_{e n'}({\bf x}_1)}{E_{\mu 0}+
E_{e 0}-E_{\mu n}-E_{e n'}}({\bf n}_1\cdot {\bf n}_2)\Psi_{\mu 0}({\bf x}_2)\Psi_{e 0}({\bf x}_1).
\end{displaymath}

In order to present an analytical estimate of this correction we transform \eqref{eq:78a} as in section II
introducing free electron Green's function:
\begin{equation}
\label{eq:78b}
\Delta b_{rec,SOPT}=\frac{16\alpha^3 M^2_eM_\mu}{m_em_\mu M_{Li}}\int d{\bf x}_3\int d{\bf x}_2\int d{\bf x}_1
\Psi^\ast_{\mu 0}({\bf x}_3)\Psi^\ast_{e 0}({\bf x}_3)\times
\end{equation}
\begin{displaymath}
\times\sum_{n\not=0}\Psi_{\mu n}({\bf x}_3)\Psi^\ast_{\mu n}({\bf x}_2)
\frac{e^{-b|{\bf x}_3-{\bf x}_1|}}{|{\bf x}_3-{\bf x}_1|}({\bf n}_1\cdot {\bf n}_2)\Psi_{\mu 0}({\bf x}_2)\Psi_{e 0}({\bf x}_1).
\end{displaymath}
After that the integration over ${\bf x}_1$ and expansion in $b$ (or in $\sqrt{M_e/M_\mu}$) give the result:
\begin{equation}
\label{eq:78c}
\int d{\bf x}_1({\bf n}_1\cdot {\bf n}_2)\frac{e^{-b|{\bf x}_3-{\bf x}_1|}}{|{\bf x}_3-{\bf x}_1|}=2\pi({\bf n}_2\cdot {\bf n}_3)
\left[\frac{4x_3}{3b}-\frac{x^2_3}{2}+\frac{2bx_3^3}{15}+\ldots\right].
\end{equation}
Taking first term in square brackets in \eqref{eq:78c} we make the angular integration and introduce the dimensionless variables
in integrals with radial wave functions:
\begin{equation}
\label{eq:78d}
\delta b_{rec,SOPT}=\nu_F\frac{64M_e}{9M_{Li}}\sqrt{\frac{M_e}{M_\mu}}\sum_{n>1}\frac{n}{\sqrt{n^2-1}}\int_0^\infty x_3^3
R_{10}(x_3)R_{n1}(x_3)dx_3\int_0^\infty x_2^2
R_{10}(x_2)R_{n1}(x_2)dx_2.
\end{equation}
Two contributions of discrete and continuous spectrum are the following:
\begin{equation}
\label{eq:78e}
\delta b^{disc}_{rec,SOPT}=\nu_F\frac{2^{11}M_e}{9M_{Li}}\sqrt{\frac{M_e}{M_\mu}}\sum_{n>1}\frac{n^6(n-1)^{2n-\frac{9}{2}}}{(n+1)^{2n+\frac{9}{2}}}=
\Biggl\{{{0.392~MHz}\atop{0.336~MHz}}.
\end{equation}
\begin{equation}
\label{eq:78f}
\delta b^{cont}_{rec,SOPT}=\nu_F\frac{2^{11}M_e}{9M_{Li}}\sqrt{\frac{M_e}{M_\mu}}\int_0^\infty\frac{ke^{-\frac{4}{k}arctg(k)}dk}{(1-e^{-2\pi/k})(k^2+1)^{3/2}}=
\Biggl\{{{0.212~MHz}\atop{0.182~MHz}}.
\end{equation}
The calculation of second term in square brackets in \eqref{eq:78c} is essentially simpler and gives the result
\begin{equation}
\label{eq:78g}
\delta b^{(2)}_{rec,SOPT}=-\nu_F\frac{2M_e^2}{3M_\mu M_{Li}}=-\Biggl\{{{0.011~MHz}\atop{0.011~MHz}}.
\end{equation}

\section{Electron vertex corrections}

\begin{figure}[htbp]
\centering
\includegraphics{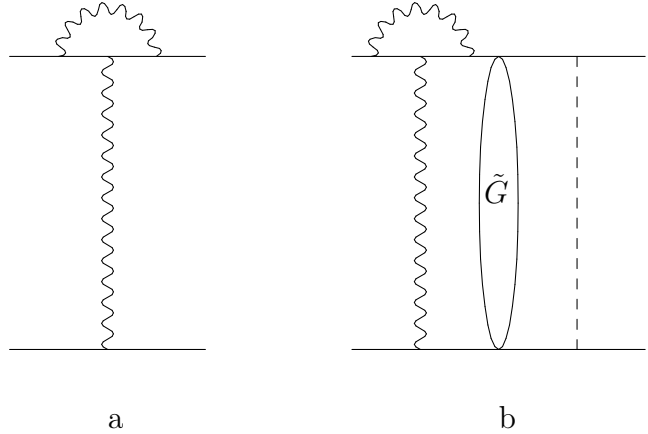}
\caption{The electron vertex corrections. The dashed line represents
the Coulomb photon. The wave line represents the hyperfine part of
the Breit potential. $\tilde G$ is the reduced Coulomb Green's
function.}
\label{fig:pic8}
\end{figure}

In the initial approximation the potential of hyperfine
splitting is determined by \eqref{eq:4}. It leads to the energy splitting
of order $\alpha^4$. In QED perturbation theory there is the
electron vertex correction to the potential \eqref{eq:4} which is defined by
the diagram in Fig.~\ref{fig:pic8}(a). In momentum representation the
corresponding operator of hyperfine interaction has the form:
\begin{equation}
\label{eq:79}
\Delta V^{hfs}_{vertex}(k^2)=-\frac{8\alpha^2}{3m_em_\mu}
\left(\frac{{\mathstrut\bm\sigma}_e{\mathstrut\bm\sigma}_\mu}{4}\right)
\left[G_M^{(e)}(k^2)-1\right],
\end{equation}
where $G_M^{(e)}(k^2)$ is the electron magnetic form factor. We
extracted for convenience the factor $\alpha/\pi$ from
$\left[G_M^{(e)}(k^2)-1\right]$.
The usual approximation for the electron magnetic form factor $G_M^{(e)}(k^2)\approx
G_M^{(e)}(0)=1+\kappa_e$ is not sufficiently accurate for the case considered here.
Indeed, characteristic momentum of the exchanged photon is $k\sim\alpha
M_\mu$. It is impossible to neglect it in the magnetic form factor
as compared with the electron mass $m_e$. So, we should use exact
one-loop expression for the magnetic form factor
\cite{t4}. Trying to improve the estimation of the
correction due to the electron anomalous magnetic moment we will use
further exact one-loop expression for the Pauli form factor $g(k^2)$
known from the QED calculation (see \cite{t4}) setting $G_M^{(e)}(k^2)-1\approx g(k^2)$.
Note that the electron form factor corresponding to the anomalous magnetic
moment is denoted frequently by $F_2$.

Using the Fourier transform of the potential \eqref{eq:79} and averaging the
obtained expression over wave functions \eqref{eq:4} we represent the
electron vertex correction to hyperfine splitting as follows:
\begin{equation}
\label{eq:80}
b_{vert,~1\gamma}=\nu_F\frac{(1+\kappa_\mu)m_e^3M_e}{81\pi^2\alpha^2 M_\mu^4}\int_0^\infty
k^2dk\left[G_M^{(e)}(k^2)-1\right]\times
\end{equation}
\begin{displaymath}
\times\left\{\left[1+\left(\frac{m_e}{6\alpha
M_\mu}\right)^2k^2\right]\left[\left(\frac{2M_e}{3M_\mu}\right)^2+\left(\frac{m_e}{6\alpha
M_\mu}\right)^2k^2\right]^2\right\}^{-1}=\Biggl\{{{40.956~MHz}\atop{40.956~MHz}},
\end{displaymath}
Let us remark that the contribution \eqref{eq:80} is of order $\alpha^5$.
Numerical value \eqref{eq:80} is obtained after numerical integration with
the one-loop expression of the electron magnetic form factor
$G_M^{(e)}(k^2)$. If we use the value $G_M^{(e)}(k^2=0)$ then the
electron vertex correction is equal $41.959$ MHz. So, using the exact
expression of the electron form factor in the one-loop
approximation we observe the 1 MHz decrease of the vertex correction
to the hyperfine splitting from $1\gamma$ interaction. Taking the
expression \eqref{eq:79} as an additional perturbation potential we have to
calculate its contribution to HFS in the second order perturbation
theory (see the diagram in Fig~\ref{fig:pic8}(b)). In this case the dashed line
represents the Coulomb Hamiltonian $\Delta H$ \eqref{eq:2}. Following the
method of the calculation formulated in previous sections we divide again total contribution from the
amplitude in Fig.~\ref{fig:pic8}(b) into two parts which correspond to the muon
ground state $(n=0)$ and muon excited intermediate states $(n\not
=0)$. In this way the first contribution with $n=0$ takes the form:
\begin{equation}
\label{eq:81}
b_{vert,~SOPT}(n=0)=\frac{8\alpha^2}{3\pi^2m_em_\mu}\int_0^\infty
k\left[G_M^{(e)}(k^2)-1\right]dk\int d{\bf x}_1\int d{\bf
x}_3\psi_{e0}({\bf x}_3)\times
\end{equation}
\begin{displaymath}
\times\Delta\tilde V_1(k,{\bf x}_3)G_e({\bf x}_1,{\bf x}_3)
V_\mu({\bf x}_1)\psi_{e0}({\bf x}_1),
\end{displaymath}
where $V_\mu({\bf x}_1)$ is defined by \eqref{eq:15} and
\begin{equation}
\label{eq:82}
\Delta \tilde V_1(k,{\bf x}_3)=\int d{\bf x}_4\psi^\ast_{\mu 0}({\bf
x}_4)\frac{\sin(k|{\bf x}_3-{\bf x}_4|)}{|{\bf x}_3-{\bf
x}_4|}\psi_{\mu 0}({\bf x}_4)=\frac{\sin\left(\frac{kx_3}{6\alpha
M_\mu}\right)}{x_3}\frac{1}{\left[1+\frac{k^2}{(6\alpha
M_\mu)^2}\right]^2}.
\end{equation}
Substituting the electron Green's function \eqref{eq:27} in \eqref{eq:82} we
transform desired relation to the integral form:
\begin{equation}
\label{eq:83}
b_{vert,~SOPT}(n=0)=\nu_F\frac{2\alpha}{81\pi^2}\left(\frac{m_e}
{\alpha M_\mu}\right)^2\left(\frac{M_e}{M_\mu}\right)^2\int_0^\infty
\frac{k\left[G_M^{(e)}(k^2)-1\right]dk}{\left[1+\frac{m_e^2k^2}{(6\alpha
M_\mu)^2}\right]^2}\times
\end{equation}
\begin{displaymath}
\times\int_0^\infty x_3e^{-\frac{2M_e}{3M_\mu}x_3}\sin\left(\frac{m_e
k}{6\alpha M_\mu}x_3\right)dx_3\int_0^\infty
x_1\left(1+\frac{x_1}{2}\right)e^{-x_1\left(1+\frac{2M_e}{3M_\mu}\right)}dx_1
\Biggl[\frac{3M_\mu}{2M_ex_>}-\ln(\frac{2M_e}{3M_\mu}x_<)-
\end{displaymath}
\begin{displaymath}
-\ln(\frac{2M_e}{3M_\mu}x_>)+Ei(\frac{2M_e}{3M_\mu}x_<)+
\frac{7}{2}-2C-\frac{M_e}{3M_\mu}(x_1+x_3)+
\frac{1-e^{\frac{2M_e}{3M_\mu}x_<}}{\frac{2M_e}{3M_\mu}x_<}\Biggr]
=\Biggl\{{{0.054~MHz}\atop{0.054~MHz}}
\end{displaymath}
All integrations over the coordinate $x_1,x_3$ are carried out
analytically and last integrations in $k$ is performed numerically.
We omit here intermediate expression before the integration in $k$ because
of its bulky form.
The second part of the vertex contribution (Fig.~\ref{fig:pic8}(b)) with $n\not =0$
can be reduced to the following form after several simplifications
which are discussed in section II:
\begin{equation}
\label{eq:84}
b_{vert,~SOPT}(n\not=0)=\nu_F\frac{27\alpha^4M_eM_\mu^3}{\pi^3}\int e^{-3\alpha M_\mu
x_2}d{\bf x}_2\int e^{-2\alpha M_e x_3}d{\bf x}_3\int e^{-3\alpha
M_\mu x_4}d{\bf x}_4\times
\end{equation}
\begin{displaymath}
\times\int_0^\infty k\sin(k|{\bf x}_3-{\bf
x}_4|)\left(G_M^{(e)}(k^2)-1\right)\frac{|{\bf x}_3-{\bf
x}_2|}{|{\bf x}_3-{\bf x}_4|}\left[\delta({\bf x}_4-{\bf x}_2)-
\psi_{\mu 0}({\bf x}_4)\psi_{\mu 0}({\bf x}_2)\right].
\end{displaymath}
We divide expression \eqref{eq:84} into two parts as provided by two terms in
square brackets of \eqref{eq:84}. After that the integration in \eqref{eq:84} over
coordinates ${\bf x}_1$ and ${\bf x}_3$ is carried out
analytically. In the issue we obtain ($\gamma_2=m_ek/6\alpha
M_\mu$):
\begin{equation}
\label{eq:85}
b_{1,vert,~SOPT}(n\not=0)=\nu_F\frac{\alpha}{162\pi^2}\left(\frac{m_e} {\alpha
M_\mu}\right)^3\frac{M_e}{M_\mu}\int_0^\infty
k^2\left[G_M^{(e)}(k^2)-1\right]dk\frac{1}{(\gamma_1^2-1)^3}\times
\end{equation}
\begin{displaymath}
\times\left[\frac{4\gamma_1(\gamma_1^2-1)}{(1+\gamma_2^2)^3}-\frac{\gamma_1
(3+\gamma_1^2)}{(1+\gamma_2^2)^2}+\frac{4\gamma_1^2(\gamma_1^2-1)}{(\gamma_1^2+
\gamma_2^2)^3}+\frac{1+3\gamma_1^2}{(\gamma_1^2+\gamma_2^2)^2}\right]=\Biggl\{{{0.472~MHz}\atop{0.472~MHz}},
\end{displaymath}
\begin{equation}
\label{eq:86}
b_{2,vert,~SOPT}(n\not=0)=-\nu_F\frac{\alpha}{162\pi^2}\left(\frac{m_e} {\alpha
M_\mu}\right)^3\frac{M_e}{M_\mu}\int_0^\infty
k^2\left[G_M^{(e)}(k^2)-1\right]dk\times
\end{equation}
\begin{displaymath}
\times\frac{1}{(1+\gamma_2^2)^2}\left[\frac{2}{(\gamma_1^2+\gamma_2^2)}-\frac{(\gamma_1+1)}
{[(1+\gamma_1)^2+\gamma_2^2]^2}-\frac{2}{(\gamma_1+1)^2+
\gamma_2^2}-\frac{\gamma_2^2-3\gamma_1^2}{(\gamma_1^2+\gamma_2^2)^3}\right]=\Biggl\{{{-0.582~MHz}\atop{-0.582~MHz}},
\end{displaymath}
It is necessary to emphasize that the theoretical error in the summary
contribution $b_{1,vert,~SOPT}(n\not =0)+b_{2,vert,~SOPT}(n\not =0)$ is determined by the
factor $\sqrt{M_e/M_\mu}$ connected with the omitted terms of the used
expansion. It can amount to $10\%$ of total results \eqref{eq:85}-\eqref{eq:86} that is the value near
0.010 MHz.

The electron vertex corrections investigated in this section have the order $\alpha^5$
in the hyperfine splitting interval. Summary value of all obtained contributions in second
order PT is equal to -0.056~MHz $(^6_3Li)$ and $(^7_3Li)$. Summing this number with the correction~\eqref{eq:80}
we obtain the value 40.900 MHz. It differs by a significant value 1.059 MHz from the
result 41.959 MHz which was obtained in the approximation of vertex correction by the electron anomalous
magnetic moment.

\begin{table}[htbp]
\caption{\label{t1}Hyperfine splitting of the ground state in
muonic lithium ions $(\mu\ e \ ^{6,7}_3Li)^+$.}
\bigskip
\label{tb1}
\begin{ruledtabular}
\begin{tabular}{|c|c|c|c|c|c|}  \hline
Contribution to the HFS & \multicolumn{2}{c|}{$b$, MHz} &\multicolumn{2}{c|}{ $c$, MHz }& Equation   \\
\hline The Fermi splitting &36140.290&36141.701 & 1674.700   & 4422.900&  \eqref{eq:10}-\eqref{eq:11} \\
\hline
Recoil correction  & -512.303&-511.012 & 8.467&22.302 &(25),(28)   \\
of order $\alpha^4(m_e/m_\mu)$  &  &   &    &   &   \\    \hline
Correction of muon anomalous &42.137  & 42.138&  --- & --- & \eqref{eq:10}  \\
magnetic moment of order $\alpha^5$& &  &  &  &   \\  \hline
Relativistic correction of order $\alpha^6$  &5.773 &5.774  & 0.535    & 1.413  & \cite{HH} \\
\hline
One-loop VP correction in $1\gamma$& 0.701   & 0.706 & 0.066    &0.175  & \eqref{eq:36},\eqref{eq:37}      \\
interaction of orders $\alpha^5,\alpha^6$  &  & & &   &  \\
\hline
One-loop VP correction  &  0.723  & 0.726 & 0.103    & 0.271 & \eqref{eq:39},\eqref{eq:40},   \\
in the second order PT&    &     &   &  &\eqref{eq:43},\eqref{eq:47},    \\
 &    &     &   &  &\eqref{eq:51},\eqref{eq:52},    \\
 &    &     &   &  &\eqref{eq:58},\eqref{eq:62},    \\
 &    &     &   &  &\eqref{eq:64}    \\   \hline
Nuclear structure correction & ---  & --- &-0.283  &  -0.707    & \eqref{eq:66}    \\
in $1\gamma$ interaction of order $\alpha^6$  &   &    &    &      &       \\ \hline
Nuclear structure correction& ---  &--- & -0.195 & -0.486    &   \eqref{eq:68}   \\
in $2\gamma$ interactions of order $\alpha^5$    &  &  &  &      &       \\
\hline
Nuclear structure correction & -0.525& -0.470 &  -0.153  &  -0.385   & \eqref{eq:71},\eqref{eq:72a},      \\
of order $\alpha^6$ in second order PT     &  &   &    &      &  \eqref{eq:74},\eqref{eq:76},     \\ \hline
Recoil correction of order & 6.430& 6.431 & --- & ---  & \eqref{eq:78} \\
$\alpha^5(m_e/m_\mu)\ln(m_e/m_\mu)$    & &  &   &   &    \\   \hline
Recoil correction of order & 0.593& 0.507 & --- & ---  & \eqref{eq:78e}-\eqref{eq:78g} \\
$\alpha^4(M_e/M_{Li})\sqrt{M_e/M_\mu}$    & &  &   &   &    \\   \hline
Electron vertex correction  &40.956  &  40.956 & ---& ---  & \eqref{eq:80}   \\
of order $\alpha^5$ in $1\gamma$ interaction&  &   & &    &    \\ \hline
Electron vertex correction &-0.056 & -0.056& ---  & --- &   \eqref{eq:83},\eqref{eq:85},  \\
of order $\alpha^5$ in second order PT&   &  &     &  & \eqref{eq:86}  \\ \hline
Summary contribution & 35724.719 &35727.401    &  1683.240  & 4445.483   &  \\
\hline
\end{tabular}
\end{ruledtabular}
\end{table}

\section{Conclusions}

In this work we have performed the analytical and numerical
calculation of hyperfine splitting intervals in muonic lithium systems $(\mu\ e\ ^{6,7}_3Li)^+$
on the basis of perturbation theory method suggested previously in the case of
muonic helium in \cite{LM}. To increase the accuracy of the calculation we take
into account several important corrections to hyperfine
splitting of the ground state of orders $\alpha^5$ and $\alpha^6$ connected with
the vacuum polarization, nuclear structure, recoil effects and
electron vertex corrections. Numerical values of different contributions to hyperfine
structure are presented in Table~\ref{tb1}.

Let us list a number of basic features of the calculation.

1. Muonic lithium atoms have a complicated hyperfine structure which appears due
to the interaction of magnetic moments of three particles. We investigate small
hyperfine splittings which can be important for experimental study.

2. In this problem there are two small parameters of fine structure constant and the ratio of
particle masses which can be used for a construction of the
perturbation interactions. Basic contributions appear in orders $\alpha^4$, $\alpha^5$ and $\alpha^6$
with the account of first and second order recoil effects.

3. The vacuum polarization effects are important in order to obtain theoretical splittings with
high accuracy. They give rise to the modification of the two-particle interaction
potential which provides the $\alpha^5\frac{M_e}{M_\mu}$-order
corrections to the hyperfine structure. We take into account the vacuum polarization corrections
in first and second orders of perturbation theory.

4. The electron vertex corrections to the coefficient $b$ should be
considered with the exact account of the one-loop magnetic
form factor of the electron because the characteristic momentum
incoming in the electron vertex operator is of order of the electron
mass.

5. Nuclear structure corrections to the ground state hyperfine
splitting are expressed in terms of electromagnetic form factors and
the charge radius of two Li nuclei.

6. Relativistic correction is obtained by means of the expression from \cite{HH}:
\begin{equation}
\label{eq:93}
\Delta\nu_{rel}=\nu_F\left(1+\frac{3}{2}(Z_1\alpha)^2-\frac{1}{3}(Z_2\alpha)^2\right),
\end{equation}
which gives contributions to both coefficients $b$ and $c$ (see Table~\ref{tb1}).

Using total numerical values of coefficients $b$ and $c$ presented
in Table~\ref{tb1} we find the following hyperfine splittings for
muonic lithium ions: $\Delta\nu_1(\mu\ e\ ^6_3Li)^+$=21572.159 MHz
and $\Delta\nu_2(\mu\ e\ ^6_3Li)^+$=14152.560 MHz;
$\Delta\nu_1(\mu\ e\ ^7_3Li)^+$=21733.056 MHz and
$\Delta\nu_2(\mu\ e\ ^7_3Li)^+$=13994.345 MHz. There is the only known to us theoretical
calculation of hyperfine structure in muonic lithium ions in
\cite{AF} which was performed by variational method and gave
hyperfine splittings: $\Delta\nu_1(\mu\ e\ ^6_3Li)^+$=21567.112 MHz
and $\Delta\nu_2(\mu\ e\ ^6_3Li)^+$=14148.678 MHz;  and
$\Delta\nu_1(\mu\ e\ ^7_3Li)^+$=21729.22 MHz
$\Delta\nu_2(\mu\ e\ ^7_3Li)^+$=13989.19 MHz.

An analysis of separate contributions to hyperfine structure coefficients $b$ and $c$ in Table~\ref{tb1}
shows that relativistic and electron vertex corrections
have large value. So, for example, the difference of our calculation from
results in \cite{AF1} for the electron vertex corrections consists the value of
order 1 MHz and relativistic corrections amount 6 MHz. The recoil contribution
from $2\gamma$ exchange amplitudes has similar value $\sim 6$ MHz.
As it follows from the calculation in \cite{AF1} only expectation values of
the delta-functions are taken into account with very high accuracy, but different corrections to the
leading order Hamiltonian were omitted.
Nonrelativistic Hamiltonian is used in \cite{AF1} and
the electron vertex corrections are taken into account in terms of anomalous magnetic
moment. So, the difference in total results between our calculation and \cite{AF1} arises
first of all from these terms. The vacuum polarization and nuclear structure corrections
influence on total result to a smaller degree. Numerical values of fundamental physical
constants in our work and \cite{AF1} coincide. It is useful also to compare our results with
\cite{AF1} in the order $\alpha^4$ with the account of electron vertex corrections in
terms of electron AMM. Such comparison shows that the difference between our work
and \cite{AF1} which lies in the region $0.7\div 1.5$ MHz for separate hyperfine splittings
remains. We consider that it is related with terms of order $\nu_F M_e^2/M_\mu^2$ which
are not taken into account exactly in our work. We included this term in total theoretical
error. Further improvement of obtained in this work results can be achieved in the
calculation of second order corrections in two small parameters $\alpha$ and $M_e/M_\mu$.

The estimate of theoretical uncertainty can be done in terms of the Fermi energy
$\nu_F$  and small parameters $\alpha$ and the ratio of the particle masses. On our
opinion, there exist several main sources of the theoretical errors.
First of all, as we mentioned above in section II the recoil corrections of order $M_e^2/M_\mu^2$
are not taken into account exactly because of a replacement of the electron Green's function by free one.
Numerically this contribution can give 0.88 MHz.
The second source of the error is related to contributions of order
$\alpha^2 \nu_F$ which appears both from QED
amplitudes and in higher orders of the perturbation theory.
In the case of two-particle bound states these corrections were calculated in \cite{BE,KP,KKS,fmms}.
Considering that they should be studied more carefully for three-particle bound states we included
a correction $\alpha^2 \nu_F\approx 1.92$ MHz in theoretical error.
Another part of theoretical error is determined by two-photon
three-body exchange amplitudes mentioned above. They are of the
fifth order over $\alpha$ and contain the recoil parameter
$(m_e/m_\alpha)\ln(m_e/m_\alpha)$, so that their possible numerical
value can be equal $\pm 0.22$ MHz.
Thereby, total theoretical uncertainty is not exceeded $\pm 2.13$
MHz. To obtain this estimate we add the above mentioned
uncertainties in quadrature.

\begin{acknowledgments}
We are grateful to A.M.~Frolov for sending us paper \cite{AF1} and useful communications.
The work is supported by the Russian Foundation for Basic Research (grant 14-02-00173)
and the Ministry of Education and Science of Russia under Competitiveness Enhancement
Program 2013-2020.
\end{acknowledgments}

\end{document}